\documentclass[twocolumn]{aastex63}
\usepackage{amsmath}
\usepackage[T1]{fontenc}
\usepackage{subfigure}
\usepackage{graphics}
\usepackage{natbib}
\usepackage{hyperref}

\usepackage{amsmath}

\newcommand{\package}[1]{\textsl{#1}}

\newcommand{\kms}{\ensuremath{\textrm{km}\,\textrm{s}^{-1}}}

\def\kmsc{~km~s$^{-1}$}

\def\hub{\ifmmode H_\circ\else H$_\circ$\fi}
\def\lsim{\mathop{\hbox{${\lower3.8pt\hbox{$<$}}\atop{\raise0.2pt\hbox{$\sim$}}$}}}
\def\gsim{\mathop{\hbox{${\lower3.8pt\hbox{$>$}}\atop{\raise0.2pt\hbox{$\sim$}}$}}}

\shorttitle {Ophiuchus Stream}
\shortauthors{Caldwell et al.}

\begin{document}

\watermark{To Appear in The Astronomical Journal}
\authorcomment1{To Appear in The Astronomical Journal}
\title{A Larger Extent for the Ophiuchus Stream}

\author{Nelson Caldwell}
\affiliation{Center for Astrophysics $|$ Harvard \& Smithsonian, 60 Garden Street, Cambridge, MA 02138, USA}

\author{Ana Bonaca}
\affiliation{Center for Astrophysics $|$ Harvard \& Smithsonian, 60 Garden Street, Cambridge, MA 02138, USA}

\author{Adrian M. Price-Whelan}
\affiliation{Department of Astrophysical Sciences, Princeton University,  Ivy Lane, Princeton, NJ 08544, USA}

\author{Branimir Sesar} 
\affiliation{Deutsche B\"orse AG, Mergenthalerallee 61, 65760 Eschborn, Germany}

\author{ Matthew G. Walker}
\affiliation{McWilliams Center for Cosmology, Department of Physics, Carnegie Mellon University, 5000 Forbes Ave., Pittsburgh, PA 15213, USA}

\correspondingauthor{Nelson Caldwell}
\email{caldwell@cfa.harvard.edu}

\shorttitle{Ophiuchus} 

\begin{abstract}
We present new kinematic data for the Ophiuchus stellar stream. Spectra have been taken of member candidates at the MMT telescope using Hectospec, Hectochelle and Binospec, which provide more than
1800 new velocities. Combined with proper motion measurements of stars in the field by the Gaia - DR2 catalog, we have derived stream membership probabilities, resulting in the detection of more than 200 likely members. These data show the stream extends to more than three times the length shown in the discovery data. A spur to the main stream is also detected. The high resolution spectra allow us to resolve the stellar velocity dispersion, found to be $1.6 \pm 0.3 $\kmsc.
\end{abstract}


\section{Introduction}\label{intro}
As discovered by \citet{bernard}, the Ophiuchus stream (OphStr) appeared to be a short and narrow stellar stream (2\fdg5\ long  by 6\arcmin \ wide), unlike most other known galactic stellar streams \citep{grill}.
\cite{bernard} used the then new PS1 catalog of stars for initial detection, and enhanced that by filtering out stars whose photometry did not fall in the area occupied by a metal-poor main sequence at heliocentric distances of $8-12$ kpc.  They calculated a luminosity for the entire stream to
be M$_{\rm v}=-3.0 \pm 0.5$, similar to that found for the lowest mass galactic globular clusters. \cite{sesar2015} followed up the imaging work with a spectroscopic project  that showed the stream to have a high heliocentric radial velocity of 290 \kmsc, nearly 300 \kms different from the mean of the majority of stars at that position. This large offset mitigates the often difficult problem of distinguishing stream members from nonmembers. Using MMT/Hectochelle and Keck/Deimos, fourteen radial velocity members were found: six blue horizontal branch stars (BHB),  two main-sequence turnoff stars, one subgiant (SG), and five red giants (RGB), four of which had spectra with high enough signal-to-noise to allow a determination of atmospheric parameters.  An analysis showed those stars to be alpha enhanced, and of low metallicity, indicating that the progenitor of the stream was likely a globular cluster.  The stream was then thought to extend from galactic longitude $l=3\fdg8\ $ to 5\fdg8\ , for a physical projected length of 300 pc at a mean heliocentric distance of 8 kpc and a variation of about 1.5 kpc over that angle, with the western end being more distant. The galactic latitude of the stream in the Sesar et al.  paper extended over the range of b$=+30.5$ to 32; the paper used the  measured velocity dispersion to estimate the initial mass to be $2 \times 10^4$ M$_\sun$.

The OphStr appeared very different from other known streams which span many tens of degrees on the sky, 
thus leading \cite{sesar2015} to postulate it was created within a 1\,Gyr short time period or required special circumstances. \cite{sesar2016} suggested that  bar-induced fanning  would have the effect of lowering the surface density at large distances along the stream, making the detection more difficult and thus causing the stream to appear shorter than it actually is.

Further spectral observations on 43 stars selected to have BHB star colors initially indicated this might be the case. Six of those stars had radial velocities indicating membership (we show below that one is not a member based on proper motions).  Several of these were positioned at large angles from the main stream,  beyond galactic longitude 6\degr \ where the discovery image shows no stream, resulting in a revised deprojected length estimate of 2 to 3 kpc, depending on which stars
were believed to be members. \cite{adrian} and \cite{hattori} investigated possible scenarios where a rotating galactic bar can create 
a shortened stream and a high velocity dispersion for nearby stars.  \cite{lane2019} used the published radial
velocities and proper motions from the recent release of the Gaia-DR2 catalog \citep[GDR2, ][]{gaia} 
to further simulate the orbit, finding that the stream is likely only on its second pericenter
passage. But there is a clear
lack of radial velocity data extensive enough to allow models to be verified, and thus a third, larger scale observational project is warranted to further explore the dimensions of the stream and the dynamical history of it. This paper describes that spectroscopic project, involving three different spectrographs on the MMT, and complemented with the GDR2 data. The spectroscopic data continues to be heterogeneous. Our paper describes the observations in the time order they were taken.

\section{New Hectospec Observations}\label{hecto}
\subsection{Target Selection}

\begin{figure}[ht]
\includegraphics[width=4.0in]{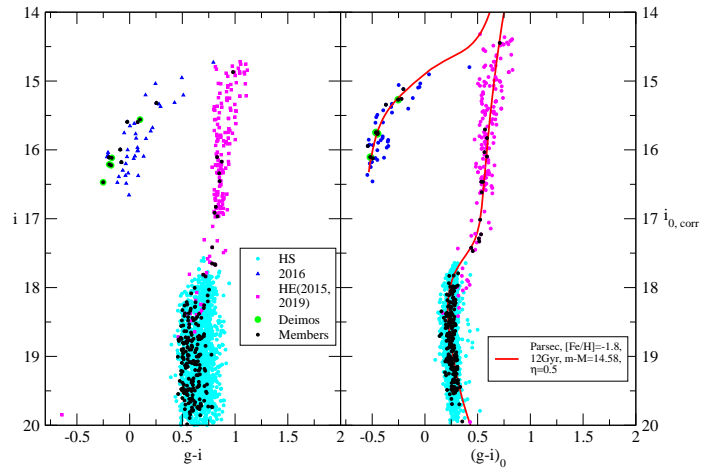} 
\caption{CMD of input stars. Left panel shows the uncorrected PS1 photometry while that on the
right has been corrected for local reddening, and the distance modulus gradient derived in
\cite{sesar2015}. Three previous sets of observed stars are indicated by different colored 
symbols, with black filled circles representing radial velocity members from all sources, mostly 
those reported here. 
Magenta circles were observed by Hectochelle (in two epochs), while green represents Deimos observations, reported
in \cite{sesar2015}. Blue circles are observations of BHB stars reported in \cite{sesar2016}. Cyan points
are the stars selected, and observed as part of the Hectospec program reported here.  A PARSEC isochone \citep{bressan}
 is also shown which has parameters: age=12\,Gyr, [Fe/H]$=-2.0$, mass loss parameter $\eta =0.5$,  and m$-$M=14.58, the mean distance modulus derived in \cite{sesar2015}.
The star selection for spectroscopic observations was based on the right panel, using MSTO stars near the isochrone as
described in the text.
A few stars appear redward of the main
group because we changed the source of the reddening map between
making the catalog and writing this paper, and some photometric values changed in the
source catalog in that period as well.
\label{fig:cmd_observed}}
\end{figure}

Although the stream is narrow in the discovery image (6\arcmin \ FWHM), it is prudent to probe a wide area to avoid missing the true distribution.  Hectospec/Hectochelle \citep{fab} with its 1\degr \ field is ideal for this, and indeed a few giant stars were observed in one field for \cite{sesar2015}, centered on the densest part of the stream. The limiting magnitude of the stars observed with Hectochelle in 2014 for  \cite{sesar2015} was {\it i}=18.8 ({\it g}=19.2), at the top of the turnoff of the main
sequence (MSTO) of the stream (see Fig \ref{fig:cmd_observed}). Most of the member stars successfully observed were on the giant branch about 1.5 mag below the level of the
horizontal branch and 1.0 mag brighter than the MSTO.
For a subsequent project, we wanted to observe 
fainter MSTO targets, which are of course far more numerous than BHB, RGB or SG stars, but would be out of reach for the high dispersion Hectochelle spectrograph (resolution=32000), and thus it was thought best to use a lower resolution option for the more extensive member search, and employ Hectospec, which gave a resolution of 2400. 

\begin{figure*}[ht]
\includegraphics[width=7.25in]{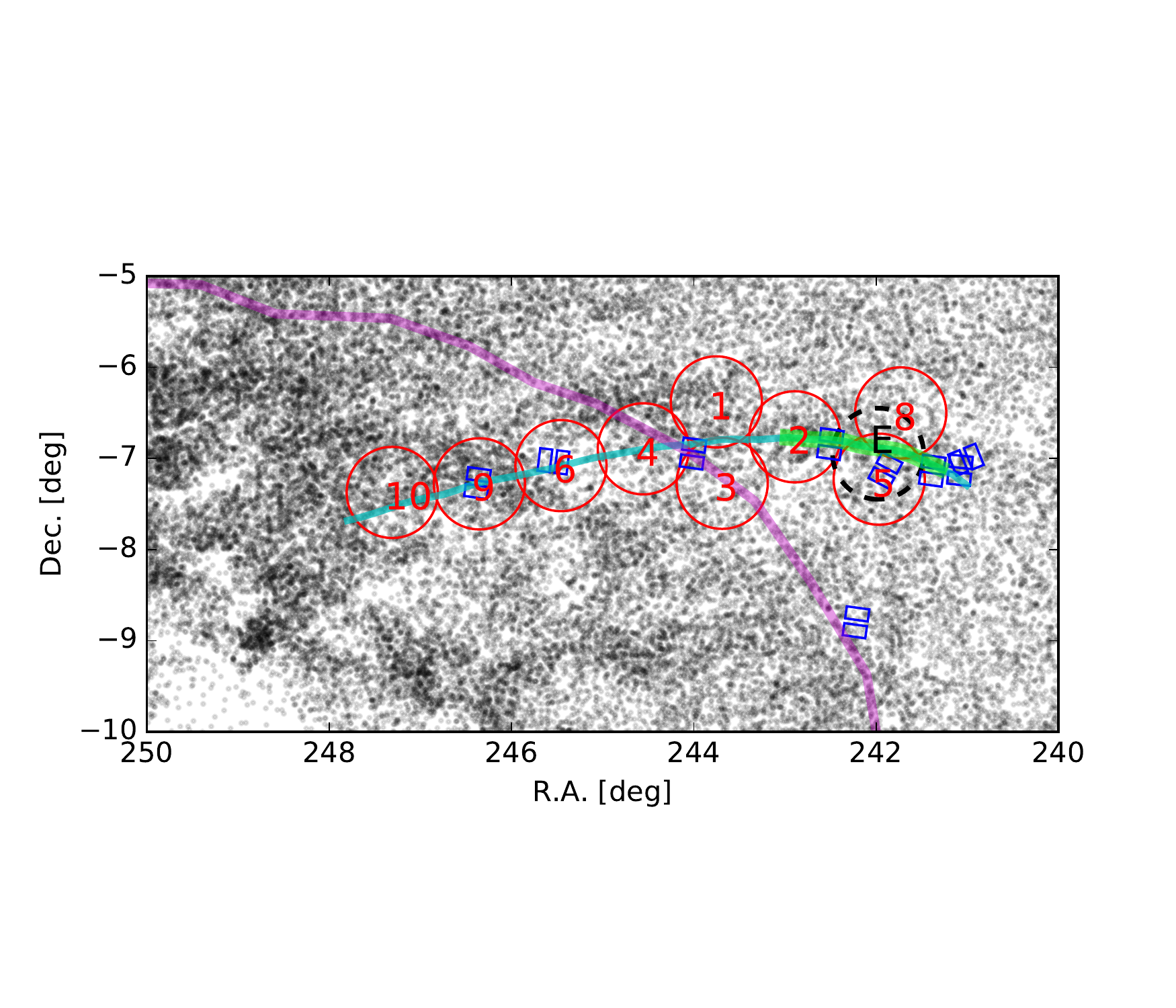} 
\caption{Distribution of stars in the Ophiuchus Stream field selected to be MSTO stars as shown in Fig \ref{fig:cmd_observed} (the cyan dots in that figure).  The nine Hectospec 1\degr \ fields are shown in red, along with the approximate distribution of the stream initially published by \cite{bernard}, shown as a thick solid green line. The positions of the circles outside of the initial stream were meant to follow the expected 
distribution determined by the orbit model of \cite{sesar2015}, shown as a thin cyan line. (Hectospec field number 7 was not observed.) The nine Binospec fields 
discussed in Section \ref{binospec} are shown 
as pairs of blue rectangles. The 2019 Hectochelle field is shown as a black dashed circle, with an ``E''. The $\zeta $ Oph \ion{H}{2} region is strong to the east (left) of 
the magenta line, and thus affects all the Hectospec fields except 1, 2, 5 and 8, with the emission getting stronger in the east direction.
\label{fig:cmd_input_dist0}}
\end{figure*}

A new observing catalog was made, using the PS1 $gri$ data \citep{kaiser}, with a data filter that selected putative MS turnoff stars. 
The original selection used PS1 data corrected for
reddening with the \cite{schlegel} dust map (the mean reddening for the stream is E(B$-$V)=0.185). Nine 1\degr \ fields were identified to sample along the model orbit of \cite{sesar2015}, where stars were included with ($g-i$)$_0$ color within
0.1 mag of the best-fit OphStr isochrone \citep[from PARSEC, corrected for extinction,][]{bressan},   bluer than  ($g-i$)$_0=0.3$ mag, and uncorrected g brighter than g = 21 mag.  \cite{sesar2015} had detected a distance modulus gradient (m-M$=14.58 - 0.2({\it l} -5$), where {\it l} is the galactic longitude),  and that was also employed in the star selection (the value is called $i_{0,corr}$). The reddening values vary by as much as 0.2 mag in E(B$-$V), so the uncorrected colors show a wider distribution than
the corrected colors. After the observations were made, we began to use the updated \citet{schlafly} map, which has a few individual star differences, and thus some stars appear outside of the original color-mag selection box in Fig \ref{fig:cmd_observed}. The net result is that selected stars had dereddened $i$ between 17 and 20 mag, and ($g-i$)$_0$ colors between 0.15 and 0.3 mag. 

Fig. \ref{fig:cmd_observed} shows the target selection in those nine fields, along with previously known members, and members found here.  Note that the members fall
within the color spread of the candidates, thus systematics have been minimized. Fig. \ref{fig:cmd_input_dist0} shows the on-sky distribution of those stars and the surrounding ones 
with the same selection criteria. 
Below, we will refine the CMD selection based on radial velocity membership for a second
round of new observations with the Binospec instrument.
The actual observing conditions varied among the different fields for the Hectospec data, so the depths explored are quite different.

The MSTO  stars are hot enough so that Balmer series lines are prominent, which will thus provide the dominant signal for radial velocity measurements.
Upon examining the raw data, we noticed a number of the fields showed Balmer emission, which we realized after the observations
were already taken was due to the eastern part of the stream overlapping  on the sky with the large $\zeta $ Oph \ion{H}{2} region \citep{haffner}. This
complicated the data reduction, as multifiber spectroscopy typically does not use local sky measurements. The fact that the dominant feature in our stars is Balmer absorption further meant
that for some spectra the emission contamination precluded the measurement of a believable stellar velocity.

\subsection{Details of Hectospec Observations}
A total of 9 separate Hectospec fields were observed in 2016 using the MMT and Hectospec
instrument with the 600gpm grating, which gave spectra in the range of $\lambda\lambda 4030-6500$\AA, 
with a resolution of 2.3\AA.
An important caveat in this data set is the variable conditions for the different
fields, making the success rates of obtaining useful spectra (which give dependable 
radial velocities) vary by a large factor among the fields. 
For six of the fields we had an average success rate of nearly 90\% but for the other three we obtained an average less than 50\%.
Sixty two stars were observed in two separate configurations.
Table \ref{obs} gives the details of the fields observed. Data were processed using the method
outlined in \citet{caldwell}
with the exception of how the sky subtraction was done. Sky fibers near the targets
were chosen rather than a sum of all sky fibers, due to the variable strength of the 
Balmer emission from the \ion{H}{2} region. This method still left a few stars poorly sky subtracted,
resulting in deep features that could be picked up by a velocity finding program (the velocity
of the \ion{H}{2} region is about $-11$ \kmsc, similar to the mean of the non-stream stars) and thus resulting
in faulty radial velocities. These spectra were noted individually and removed from the catalog. More than 1700 stars were observed with Hectospec.

\begin{deluxetable*}{llrrrlrr}
\tablecolumns{8}
\tablewidth{0pc}
\tablecaption{Observation Log \label{obs} }
\tablehead{\colhead{field} & \colhead{Inst\tablenotemark{a}} & \colhead{Date} &\colhead{RA} & \colhead{Dec}  & \colhead{Exposures} &   \colhead{Seeing\tablenotemark{b}} &\colhead{\% Good} \\
\colhead{} & \colhead{} & \colhead{} & \multicolumn{2}{c}{J2000}  & \colhead{(sec)} &   \colhead{} &\colhead{}  \\
}
\startdata
1 &H& 2016.0311 &  16:15:00.7 &    -06:22:46.9 &$3\times1800$ &1.0 &68\\
2 &H& 2016.0313 &  16:11:33.9 &    -06:45:45.9 &$3\times1500$ &0.7 &94    \\
3 &H& 2016.0314 &  16:14:45.3 &    -07:16:22.9 &$4\times1800$ &0.9 &98  \\  
4 &H& 2016.0316 &  16:18:12.9 &    -06:53:40.9 &$3\times1800$ &0.7 &93  \\
5 &H& 2016.0417 &  16:07:51.7 &    -07:13:40.6 &$3\times1800$ &2.2 &31 \\
6 &H& 2016.0417 &  16:21:49.8 &    -07:04:44.2 &$3\times2100$ &2.2 &47  \\
8 &H& 2016.0418 &  16:06:55.1 &    -06:30:03.6 &$3\times2100$ &1.7 &83 \\
9 &H& 2016.0419 &  16:25:24.8 &    -07:16:44.4 &$4\times1800$ &1.5 &44   \\
10 &H&2016.0317 &  16:29:14.1 &    -07:22:22.9 &$3\times1800$ &0.8 &91\\                                                          
11 &B&2018.0410 &  16:07:35.4 &    -07:07:52.5 &$3\times600$ &  & 81\\                                                          
12 &B&2018.0410 &  16:04:03.5 &    -07:00:53.1 &$3\times600$ &  &82\\                                                          
13 &B&2018.0411 &  16:08:53.0 &    -08:47:53.2 &$3\times600$ & & 98\\                                                          
14 &B&2018.0411 &  16:22:09.8 &    -07:01:52.1 &$3\times600$ & &91\\                                                          
15 &B&2018.0604 &  16:05:32.1 &    -07:08:04.9 &$3\times1200$ & &88\\                                                          
16 &B&2018.0605 &  16:04:19.2 &    -07:07:47.5 &$3\times1200$ &&79\\                                                          
17 &B&2018.0608 &  16:10:00.4 &    -06:50:34.7 &$3\times1200$ &&95\\                                                          
18 &B&2018.0612 &  16:16:02.3 &    -06:57:45.9 &$3\times1200$ &&72\\                                                          
19 &B&2018.0613 &  16:25:29.7 &    -07:16:13.3 &$3\times1200$ &&89\\                                                          
20 &E&2019.0622 &  16:07:54.8 &    -06:56:54.0 &$6\times2600$ & 1.1 &43\\                                                          
\enddata
\tablenotetext{a}{H=Hectospec, B=Binospec, E=Hectochelle}
\tablenotetext{b}{in arcsecs. Seeing values were not recorded during the Binospec runs.}
\end{deluxetable*}

Heliocentric radial velocities were found using {\it xcsao} \citep{kurtz}, in the spectral range of $\lambda\lambda 4200-5300$\AA; this range gave the cleanest cross-correlation functions and hence 
the lowest formal errors. Templates used were model atmosphere spectra supplied by C. 
Johnson, all at T$_{eff}=6000$ and log $g=4$, but four metallicities ([Fe/H] $= -2, -1.5, -1.0, -0.5$ and $0$). The general features of these templates match those of
the expected spectra of the targeted MSTO stars (the predominant lines are the Balmer series, though Mgb and the stronger Fe lines are also present).

For many of the repeated stars, one observation had S/N too low to obtain a significant velocity; just five repeated stars had small
enough errors for analysis. For those pairs we determine an RMS difference of 10 \kmsc, meaning the true RMS is $10/\sqrt{2}= 7$ \kmsc,
 somewhat smaller than the mean formal velocity uncertainty for the stars
of 13 \kmsc. Four stars were in common with the earlier Hectochelle data, where the formal velocity uncertainties are an order
of magnitude smaller. We find the RMS difference between these to be 10 \kmsc, with the formal RMS being 20 \kmsc. Again, the true RMS is likely to be smaller than the formal errors reported in
the Table \ref{spectra} (below).

It would be ideal to do a full fit to these spectra to derive stellar properties,
but we postpone that till a later paper, so that we can present the main
result in a timely manner. The high radial velocity of OphStr made it unlikely we
would have included too many nonmembers by simply making an initial membership 
selection using only radial velocity.

\subsection{The Stream in Radial Velocity Space}

\begin{figure}[ht!]
\includegraphics[width=4.0in]{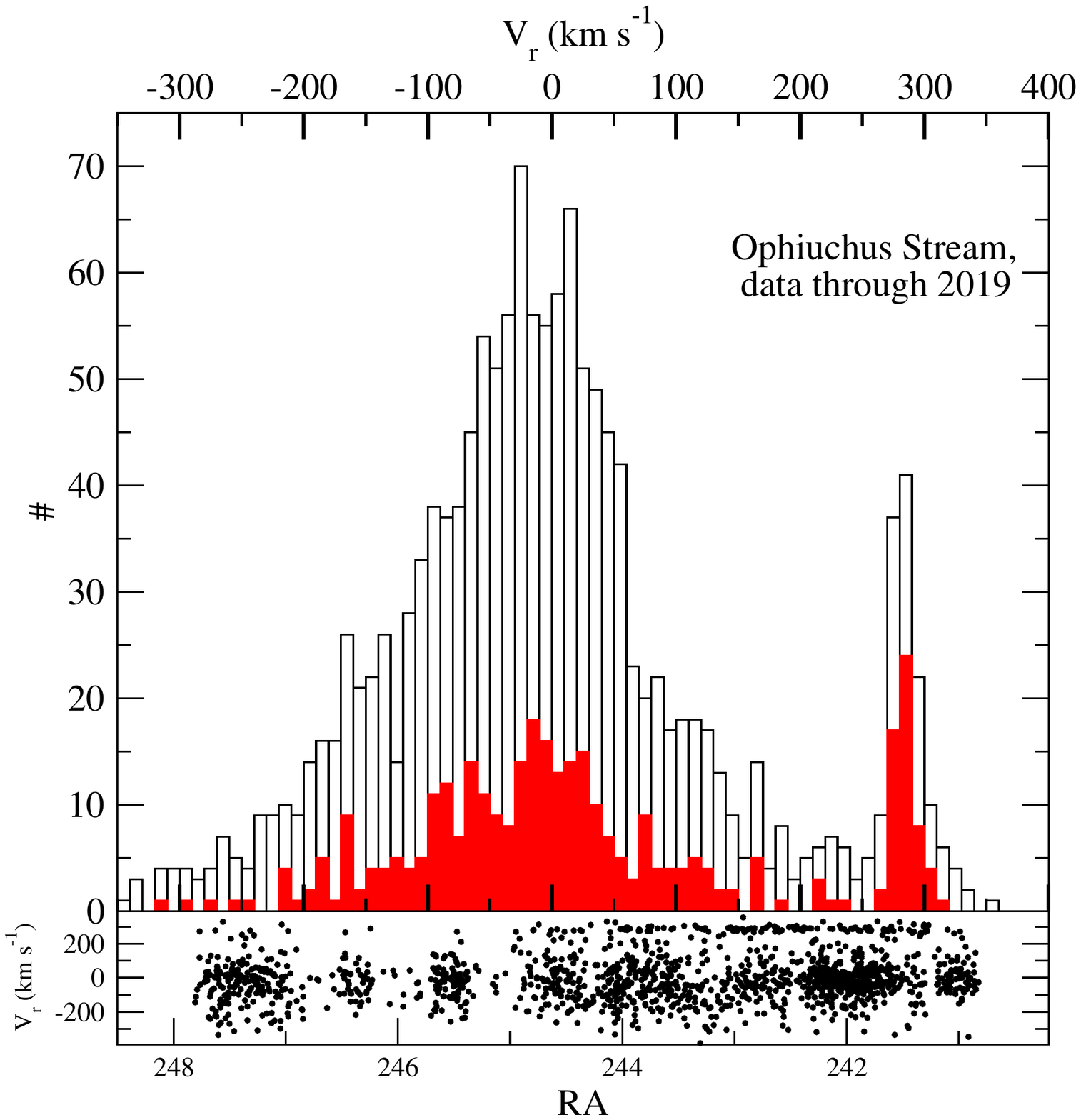} 
\caption{Top: The histogram of well-determined radial velocities from Hectospec, the two previous papers \citep{sesar2015, sesar2016}, the Binospec data, and the Hectochelle data from this paper, binned into 10 \kms bins. The stream stands out clearly at a velocity near $+300$\kmsc.  The red histogram shows just stars within 5\arcmin \ of the orbit shown in  Fig. \ref{fig:cmd_input_dist0}. Bottom: the distribution of
radial velocities with RA, also showing the stream separation in velocity.
\label{fig:vel}}
\end{figure}

We were fortunate that the OphStr has a radial velocity of nearly 300 \kms different from the mean
of the field stars present. This allows us to make a fairly clean initial assessment of 
membership based on radial velocity alone, which we will then use to refine the CMD selection and
include a proper motion and position selection as well.

We present the histogram of all radial velocities with significant detections in Fig \ref{fig:vel}. Included here
are the new Hectospec data, the radial velocities from \cite{sesar2015} and \cite{sesar2016}, as
well as the other new data to be reported in sections below. The mean velocity is around 290 \kmsc, which we will refine below. Stars within 5\arcmin \ of the \cite{sesar2015} orbit model (in red) show a stronger velocity contrast than
the sample as a whole.
 The bottom panel, plotting radial velocities as a function of RA, also clearly shows the separation of the stream from the background.
Table \ref{spectra} contains the spectroscopic data. Items listed are those for which we have spectra of any quality. If the derived radial velocity was considered flawed, the velocity entry is blank, but the correlation coefficient R is listed (in broad terms, R values greater than 4 means that 
the derived radial velocity can be trusted.)  Items where no GDR2 proper motions are available are still listed if a spectrum was taken, even if it was poor. The table is organized such that stars with membership
probabilities greater than zero are listed first, in order of RA. The second group has stars with
probabilities equal to zero, or for which we have no determination.

\begin{deluxetable*}{lrrrrrrlrrrrrr}
\rotate
\tablewidth{9.0truein}
\hoffset=-0.9truein
\tablecolumns{14}
\tabletypesize{\tiny}
\tablecaption{Spectroscopic Observations\label{spectra}. This is a stub; full table to appear in the journal.}
\tablehead{
\colhead{S\tablenotemark{a}} &\colhead{RA} &\colhead{DEC} &\colhead{g\tablenotemark{b}}  &\colhead{i} &\colhead{($g-i$)$_0$\tablenotemark{c}}  &\colhead{i$_0$} & \colhead{V}  &\colhead{R} &\colhead{$\mu_{\alpha}^{*}$} &\colhead{$\mu_{\delta}$} &\colhead{$\pi$\tablenotemark{d}}   & \colhead{D\tablenotemark{e}}  & \colhead{P$_{\rm mem}$}\\
\colhead{} & \multicolumn{2}{c}{J2000} &\colhead{} &\colhead{} &\colhead{}&\colhead{}&\colhead{\kms}  &\colhead{} &\colhead{mas~yr$^{-1}$ } &\colhead{mas~yr$^{-1}$}  &\colhead{mas}  &\colhead{arcmin} & \colhead{} \\
}
\startdata
B&	240.97155&	-7.18619&	19.515$\pm	0.027$&	19.010$\pm	0.021$&	 0.244$\pm	 0.034$&	18.715&	 234.1$\pm	20.4$&	   5.6&	-3.948$\pm	 0.934$&	-4.909$\pm	 0.490$&	 0.308$\pm	 0.424$&	  9.7&	0.93	\\
B&	240.98628&	-7.02181&	19.462$\pm	0.015$&	19.005$\pm	0.035$&	 0.195$\pm	 0.038$&	18.710&	 277.9$\pm	50.3$&	   3.0&	-3.562$\pm	 0.979$&	-3.148$\pm	 0.477$&	 0.112$\pm	 0.444$&	  0.2&	0.98	\\
B&	240.98870&	-7.18613&	19.885$\pm	0.042$&	19.383$\pm	0.014$&	 0.239$\pm	 0.044$&	19.087&	 272.0$\pm	19.7$&	   5.2&	-3.047$\pm	 1.786$&	-3.562$\pm	 0.883$&	-0.580$\pm	 0.724$&	  9.9&	0.97	\\
B&	240.99936&	-6.89888&	20.006$\pm	0.054$&	19.546$\pm	0.041$&	 0.204$\pm	 0.068$&	19.258&	 206.9$\pm	28.8$&	   3.1&	-5.764$\pm	 1.248$&	-4.237$\pm	 0.659$&	-0.023$\pm	 0.677$&	  6.9&	0.70	\\
B&	241.04199&	-7.18313&	19.337$\pm	0.029$&	18.792$\pm	0.025$&	 0.281$\pm	 0.038$&	18.494&	 260.4$\pm	17.1$&	   7.0&	-4.118$\pm	 0.922$&	-5.120$\pm	 0.423$&	-0.216$\pm	 0.382$&	 10.4&	0.98	\\
B,B&	241.09955&	-7.08248&	19.402$\pm	0.027$&	18.892$\pm	0.023$&	 0.243$\pm	 0.035$&	18.591&	 282.0$\pm	15.5$&	   5.3&	-3.974$\pm	 0.858$&	-4.302$\pm	 0.414$&	-0.195$\pm	 0.380$&	  5.1&	0.99	\\
B&	241.10806&	-7.14182&	21.102$\pm	0.088$&	20.480$\pm	0.099$&	 0.354$\pm	 0.132$&	20.178&	 264.0$\pm	31.8$&	   3.3&	-0.052$\pm	 4.092$&	-4.078$\pm	 1.533$&	-0.418$\pm	 1.443$&	  8.7&	0.91	\\
B,B&	241.11475&	-7.17195&	20.105$\pm	0.038$&	19.605$\pm	0.051$&	 0.234$\pm	 0.064$&	19.305&	 286.5$\pm	22.0$&	   2.8&	-4.217$\pm	 1.659$&	-4.363$\pm	 0.790$&	 0.307$\pm	 0.785$&	 10.5&	0.99	\\
B&	241.12453&	-7.16020&	20.464$\pm	0.054$&	19.895$\pm	0.043$&	 0.302$\pm	 0.069$&	19.594&	 353.5$\pm	24.3$&	   3.5&	-8.540$\pm	 1.860$&	-4.556$\pm	 0.847$&	 1.073$\pm	 0.861$&	 10.0&	0.89	\\
B&	241.15885&	-7.08421&	20.049$\pm	0.087$&	19.503$\pm	0.044$&	 0.280$\pm	 0.097$&	19.204&	 337.6$\pm	23.6$&	   3.7&	-4.732$\pm	 1.559$&	-4.689$\pm	 0.671$&	 0.450$\pm	 0.702$&	  5.8&	0.99	\\
B&	241.18693&	-7.09031&	20.040$\pm	0.077$&	19.481$\pm	0.034$&	 0.293$\pm	 0.084$&	19.181&	 307.8$\pm	25.7$&	   3.5&	-6.119$\pm	 1.481$&	-4.754$\pm	 0.630$&	 0.887$\pm	 0.741$&	  6.5&	0.99	\\
B&	241.25540&	-7.11738&	20.229$\pm	0.046$&	19.701$\pm	0.076$&	 0.265$\pm	 0.089$&	19.403&	 291.5$\pm	20.7$&	   5.0&	-0.449$\pm	 1.558$&	-5.054$\pm	 0.712$&	-0.704$\pm	 0.868$&	  8.8&	0.93	\\
B&	241.26994&	-7.09292&	19.700$\pm	0.029$&	19.211$\pm	0.038$&	 0.227$\pm	 0.048$&	18.914&	 308.6$\pm	22.5$&	   5.5&	-5.920$\pm	 1.036$&	-4.117$\pm	 0.492$&	 0.679$\pm	 0.564$&	  7.5&	0.99	\\
B&	241.28172&	-7.08352&	19.034$\pm	0.033$&	18.479$\pm	0.016$&	 0.293$\pm	 0.037$&	18.183&	 276.4$\pm	15.8$&	   7.4&	-4.146$\pm	 0.634$&	-4.601$\pm	 0.306$&	 0.014$\pm	 0.346$&	  7.1&	0.99	\\
B&	241.30041&	-7.09073&	19.995$\pm	0.039$&	19.497$\pm	0.033$&	 0.237$\pm	 0.051$&	19.201&	 269.1$\pm	17.0$&	   6.6&	-3.246$\pm	 1.286$&	-4.573$\pm	 0.717$&	-0.271$\pm	 0.567$&	  7.7&	0.98	\\
B&	241.30960&	-7.02269&	19.286$\pm	0.036$&	18.740$\pm	0.027$&	 0.282$\pm	 0.045$&	18.441&	 311.4$\pm	18.0$&	   6.5&	-6.173$\pm	 0.802$&	-1.888$\pm	 0.363$&	 0.134$\pm	 0.327$&	  3.8&	0.55	\\
B&	241.31232&	-7.21352&	19.393$\pm	0.032$&	18.899$\pm	0.016$&	 0.230$\pm	 0.036$&	18.602&	 301.4$\pm	15.9$&	   6.9&	-3.796$\pm	 0.789$&	-4.576$\pm	 0.385$&	 0.484$\pm	 0.519$&	 15.1&	0.99	\\
B&	241.33822&	-7.06355&	19.238$\pm	0.017$&	18.746$\pm	0.023$&	 0.231$\pm	 0.029$&	18.450&	 256.0$\pm	21.4$&	   6.0&	-4.761$\pm	 0.746$&	-4.041$\pm	 0.355$&	 0.053$\pm	 0.323$&	  6.5&	0.98	\\
B&	241.33883&	-7.10237&	19.058$\pm	0.036$&	18.550$\pm	0.031$&	 0.246$\pm	 0.048$&	18.254&	 263.3$\pm	14.8$&	   7.2&	-4.824$\pm	 0.626$&	-4.591$\pm	 0.305$&	-0.036$\pm	 0.282$&	  8.8&	0.99	\\
B&	241.34530&	-7.08610&	19.925$\pm	0.071$&	19.410$\pm	0.029$&	 0.253$\pm	 0.077$&	19.115&	 244.0$\pm	14.4$&	   8.4&	-4.043$\pm	 1.332$&	-5.079$\pm	 0.739$&	 0.470$\pm	 0.577$&	  7.9&	0.95	\\
B&	241.39887&	-7.02366&	19.832$\pm	0.073$&	19.356$\pm	0.046$&	 0.209$\pm	 0.086$&	19.056&	 270.7$\pm	21.6$&	   5.2&	-3.421$\pm	 1.463$&	-4.219$\pm	 0.645$&	-0.995$\pm	 0.546$&	  4.7&	0.99	\\
H&	241.40146&	-6.36448&	19.730$\pm	0.054$&	19.112$\pm	0.042$&	 0.277$\pm	 0.068$&	18.727&	 313.9$\pm	28.4$&	   5.9&	-4.459$\pm	 0.931$&	-3.615$\pm	 0.542$&	-0.195$\pm	 0.443$&	 34.4&	0.97	\\
B&	241.41009&	-7.05488&	20.419$\pm	0.061$&	19.844$\pm	0.059$&	 0.310$\pm	 0.085$&	19.544&	 331.8$\pm	26.9$&	   4.3&	-5.694$\pm	 2.218$&	-3.673$\pm	 0.963$&	-0.606$\pm	 0.815$&	  6.6&	0.98	\\
B&	241.41503&	-7.07013&	20.042$\pm	0.043$&	19.475$\pm	0.086$&	 0.301$\pm	 0.096$&	19.175&	 263.4$\pm	12.8$&	   7.5&	-5.383$\pm	 1.536$&	-5.267$\pm	 0.714$&	 1.295$\pm	 0.631$&	  7.6&	0.98	\\
B&	241.43110&	-7.06141&	19.964$\pm	0.056$&	19.475$\pm	0.049$&	 0.222$\pm	 0.074$&	19.174&	 274.3$\pm	23.4$&	   4.7&	-6.056$\pm	 1.488$&	-3.814$\pm	 0.744$&	 0.447$\pm	 0.608$&	  7.2&	0.98	\\
B&	241.44318&	-7.01940&	20.047$\pm	0.073$&	19.531$\pm	0.032$&	 0.247$\pm	 0.080$&	19.229&	 280.2$\pm	16.5$&	   7.4&	-4.491$\pm	 1.518$&	-4.615$\pm	 0.835$&	 0.988$\pm	 0.644$&	  4.9&	0.99	\\
B&	241.44472&	-6.99315&	19.964$\pm	0.035$&	19.459$\pm	0.051$&	 0.237$\pm	 0.062$&	19.155&	 275.7$\pm	21.9$&	   5.0&	-5.604$\pm	 1.142$&	-4.136$\pm	 0.611$&	-0.526$\pm	 0.525$&	  3.3&	0.99	\\
B&	241.44686&	-7.02205&	20.161$\pm	0.076$&	19.641$\pm	0.091$&	 0.252$\pm	 0.119$&	19.338&	 220.6$\pm	22.3$&	   4.8&	-3.877$\pm	 1.440$&	-4.332$\pm	 0.768$&	-0.201$\pm	 0.714$&	  5.0&	0.85	\\
B&	241.47695&	-6.98119&	20.182$\pm	0.074$&	19.618$\pm	0.061$&	 0.295$\pm	 0.096$&	19.313&	 300.1$\pm	21.7$&	   4.2&	-3.504$\pm	 1.304$&	-4.792$\pm	 0.924$&	-0.561$\pm	 0.706$&	  2.9&	0.99	\\
D&	241.49994&	-7.03410&	16.018$\pm	0.021$&	16.211$\pm	0.013$&	-0.465$\pm	 0.025$&	15.903&	 285.3$\pm	 1.9$&	   5.0&	-4.620$\pm	 0.197$&	-4.382$\pm	 0.120$&	 0.040$\pm	 0.096$&	  6.2&	0.99	\\

\enddata
\tablenotetext{a}{Source. A=\cite{sesar2016}, E=Hectochelle\citep{sesar2015}, D=Deimos \citep{sesar2015}, H=Hectospec, B=Binospec, F=Hectochelle 2019 (the latter three reported here)}
\tablenotetext{b}{Photometry from PS1}
\tablenotetext{c}{Dereddened using the maps of \cite{schlafly}}
\tablenotetext{d}{Parallaxes are as reported in GDR2; we do not add in the $+0.05$ mas
correction as discussed in the text.}
\tablenotetext{e}{Angular distance to stream fiducial}
\end{deluxetable*}

\subsubsection{Reddening}
We used the \cite{schlafly} map to correct the PS1 photometry, noting that  the kernel size for this map is about the width of the stream, which could lead to some circular problems in
determining the reddenings. The mean reddening over the stream is 0.185, with a range from 0.408 to 0.139 (the median value is similar to the mean).

\subsubsection{Distance confirmation}\label{cmd_distance}

\begin{figure}[ht]
\includegraphics[width=4.0in]{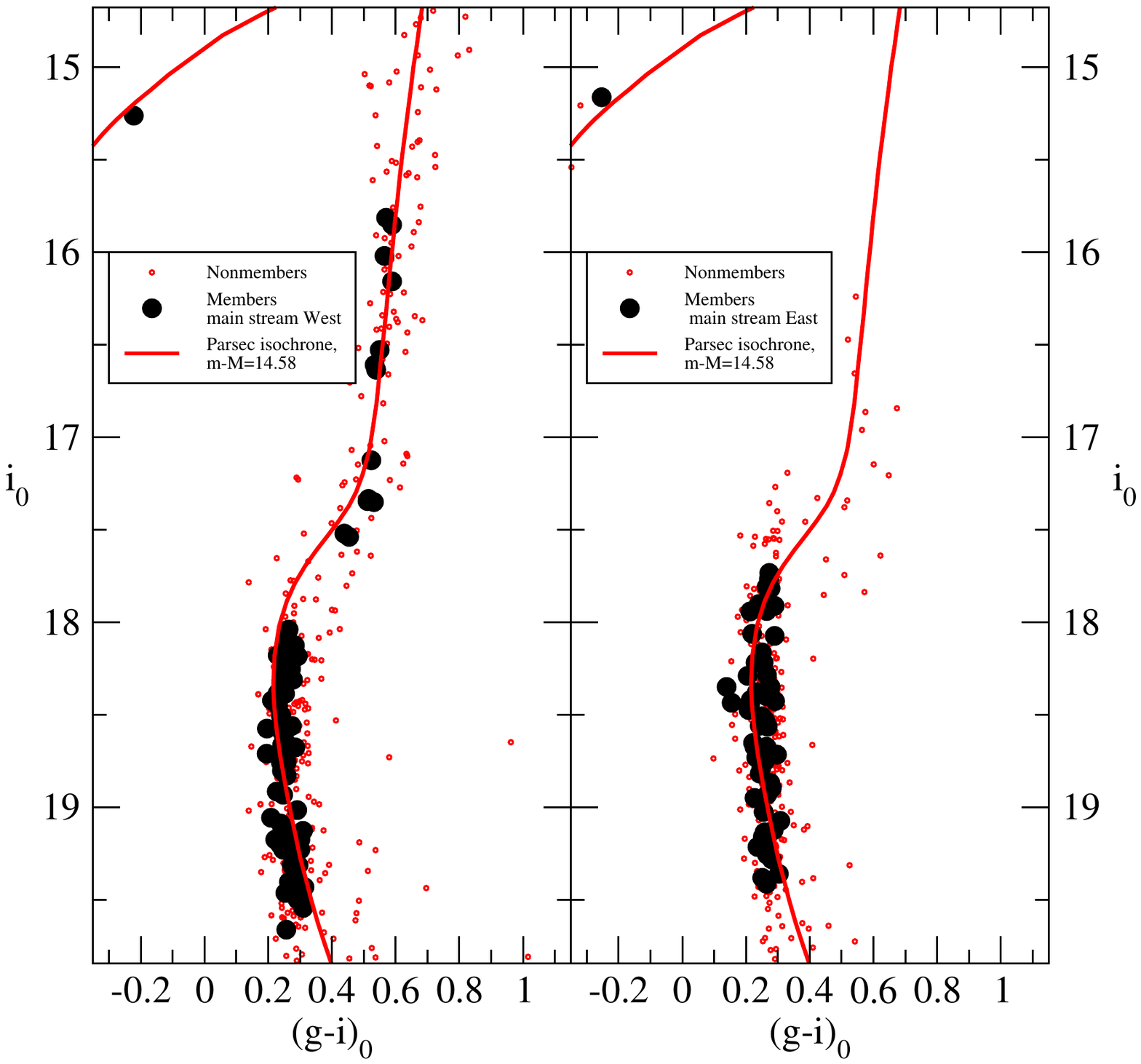} 
\caption{ CMD of high probability member stars (P$_{\rm mem} > 0.9$), divided into two parts at opposite ends of the densest part of the stream, to show the evidence for a distance difference for those parts. Left: the western part, from ${\it l}$ = 3\fdg7\ to 5\fdg4\ (75 member stars). Right: the eastern part, from ${\it l}$=5\fdg4\ to 7\fdg0\ (55 stars).
The Parsec isochrone uses the same distance modulus in both panels. 
Note the MS turnoff occurs at a brighter
magnitude for the eastern part of the stream. These stars are within 10\arcmin \ perpendicular to the long axis of the stream. Red circles indicate stars observed but which had
radial velocities precluding membership.
\label{fig:dm_grad}}
\end{figure}

From their 14 member stars, \cite{sesar2015} derived a distance modulus and gradient in the modulus, specifically
m-M$=14.58 - 0.2({\it l}-5$), where {\it l} is the galactic longitude (thus giving a distance
of 8.2 kpc at ${\it l}=5$). We confirmed this in
general for our new sample of MSTO stars we have observed, 
without rederiving a new value with
the rigor that \cite{sesar2015} employed. Fig. \ref{fig:dm_grad} shows the CMD
of stars with membership probability P$_{\rm mem} > 0.9$ (see Sect. \ref{em}) in two widely separated parts of the stream: the western part, from
${\it l}$ = 3\fdg7\ to 5\fdg4\ (75 member stars), and the eastern part,

from ${\it l}$=5\fdg4\ to 7\fdg0\ (55 stars).
A simple comparison of the PS1 ($g-i$) CMD of the similar metallicity globular cluster
NGC~5466 with that of the OphStr members (using dereddened photometry) 
reveals that the stated distance modulus and gradient is
satisfactory for the MSTO stars and the HB stars to within 0.15 mag, 
but is not as good for the
member RGB stars (11 in total, including the ones found in the Hectochelle
observations), with the stars being 0.05 mag bluer in the ($g-i$) color than expected from the
relative position of the MSTO stars, which could be due to a difference in stellar
populations between NGC~5466 and the stream.

\section{Binospec Data}\label{binospec}
This instrument is a new low-resolution, multi-slit spectrograph on the MMT, which features two $8\times 15$ \arcmin \ fields \citep{fab2}. Although the field is smaller than Hectospec, the extra depth afforded
made it attractive, and some telescope time became available during the first months of operation.
We 
employed the 1000 gpm grating which gives spectral resolution of 1.5\AA \ over a range of 1500\AA. We centered the spectra at 5200\AA \ which allowed the inclusion of the H$\beta$ line as well as the Mg~b lines. Slit widths were 1\farcs, and lengths were typically more than 20\farcs \ long. The OphStr Binospec observation information is also contained in Table \ref{obs}.

We selected fields to overlap some of the Hecto fields, particularly where the Hecto
data was compromised by bad observing conditions, but also to explore the ends of the apparent
stream, as well as a field well off of the visual stream as a control, as shown in Fig. \ref{fig:cmd_input_dist0}.

Stars on the main sequence as described above were selected  (they generally 
occupy the same part of the CMD as the Hectospec observations shown in Fig \ref{fig:cmd_observed}) , as well as some
lower RGB stars (none of the latter turned out to be members). The catalog had fainter stars
than the Hectospec sample (down to $i=21$), but only 20\% of the spectra for stars fainter
than $i=21$ provided good velocities (furthermore, only a handful of stars fainter than
$i=20.4$ have GDR2 measurements). Two separate runs were obtained in 2018, with the first extending to $i=21$, while
the second only to i$=$20. The statistics reported in Table \ref{obs} limit the first set to
$i=20$ for comparison. Heliocentric radial velocities were found
as above for the Hectospec data, using the same templates. More than 800 stars were observed, making for a total of 2600 unique objects observed in all the various projects to date to obtain
spectra in OphStr. Six stars were observed twice, giving an RMS error for measurement of 10\kmsc.

There were 26 stars observed both with Hectospec and Binospec; 23 of those have good velocities for both
measurements. The mean velocity difference of those was +1.1\kms (Hectospec $-$ Binospec), with an RMS 
of the differences of 16\kmsc, or 11\kms per star in the mean for each sample if they have the same error distribution.
The mean formal errors provided by xcsao for those Binospec stars is 15\kmsc, so the formal errors are somewhat
larger than the true errors, but we will assume the formal errors are accurate for the analyses below.

\section{Using Gaia DR2 to Refine a Candidate List}\label{gaia}

\begin{figure}[ht]
\plotone{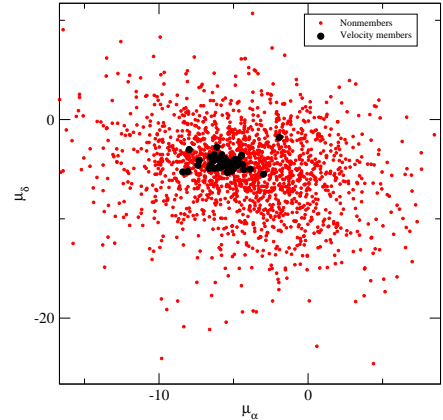} 
\caption{The distribution of Gaia DR2 proper motions for OphStr stars observed
spectroscopically with Hectospec, with RA between 241 and 243\degr, the densest part of the stream. 
Those stars with radial velocities  $260 < $V$ < 318$ \kms are shown 
in dark, larger symbols here. The mean proper motion is clearly identified in this plot. Radial velocity member stars with errant proper motions are excluded in the more exacting membership determination provided by the EM method.
\label{fig:pm_both}}
\end{figure}

We next incorporated the 
new Gaia DR2 proper motion catalogs \citep{gaia}, which allowed us to identify additional member candidates, leading up
to a second round of spectroscopic observations with the high resolution Hectochelle instrument
on the MMT.
We determined the general proper motion of the stream by examining the proper
motions of the members selected purely on previously determined
radial velocities, specifically those with $260 < $V$ < 318$ \kmsc.  
As OphStr extends over several degrees in RA (and galactic longitude), there may be a gradient in proper motions
\citep[as reported in][]{sesar2015}. 
We first looked at the mean values for the apparently densest part, between RA=241 and 243\degr . We included
proper motion data only where the errors were less than 1.8 mas yr$^{-1}$ for $\mu_{\alpha}$ and less
than 1.04 mas yr$^{-1}$ for  $\mu_{\delta}$, both of which excluded the worst 10\% of the data. These restrictions left us with 45 stars whose radial velocities indicated membership.
Fig \ref{fig:pm_both} shows this distribution. As expected, the members have a small range in proper motion, 
while the field stars have a large dispersion, though the mean values are similar for both groups.

\begin{figure}[ht]
\includegraphics[width=4.0in]{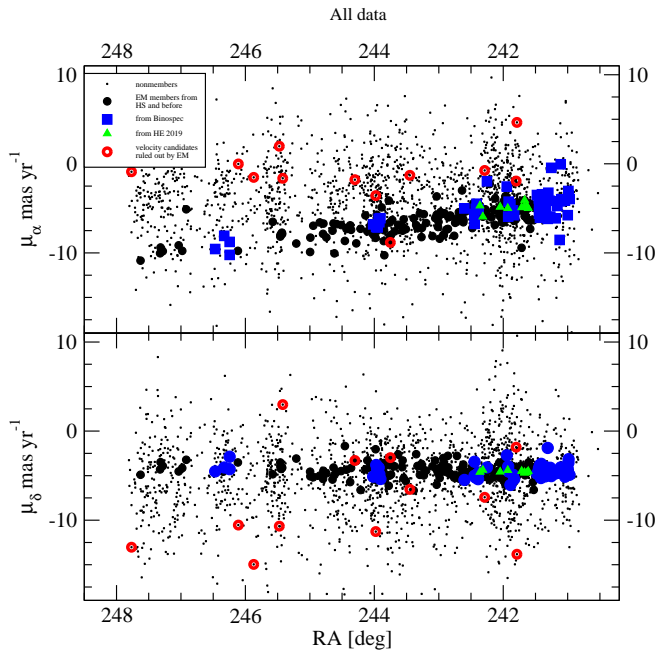} 
\caption{The relation of proper motion in both coordinates with RA for OphStr stars observed
spectroscopically with the various instruments reported here.
Those stars with radial velocities  $260 < $V$ < 318$ \kms are shown 
in larger symbols here, in several colors. Small symbols are spectroscopic observations
of non-member stars. The EM technique used these clear relations along 
with the radial velocities to classify
the outlier stars (shown in red) as non-members. Large black symbols are members
identified using Hectospec and earlier data, while blue represents members found in
Binospec data, and green was used for Hectochelle members.
The patchiness
in this diagram is due to the discrete nature of the observations (we used fields that do
not always overlap).
\label{fig:oph_pm8}}
\end{figure}

\cite{sesar2015} measured proper motions 
of 14 stars determined from astrometry in the USNO-B, 2MASS and PS1 catalogs. 
These agree in general with the GDR2 values \citep[once transformed to the galactic values
that][published]{sesar2015} with the exception of one star (star ``rgb4'') for which the
measured values are very different from the other radial velocity members in his list, likely due
to stated problems with blending. The GDR2 values for that star
are similar to the 
other velocity members, thus this star is considered a member here.  
 For all the stars in common, there is a systematic difference of $+1.0$ mas yr$^{-1}$ in both  $\mu_{l}$ and  $\mu_{b}$, with an
RMS of the differences of 1 mas yr$^{-1}$, once we exclude the one bad star. \cite{lane2019} also
found the systematic error in the \cite{sesar2015} proper motions. Our analysis of the proper
motions in galactic coordinates shows agreement with the \cite{lane2019} description of the proper motions.

Next we derived the relation of  $\mu_{\alpha}$ and  $\mu_{\delta}$ with RA (galactic longitude would work as well). This is shown in Fig. \ref{fig:oph_pm8} (all of the large points,
no matter the color). We then selected stars as candidates for
membership by insisting that the stars in the general area of the known stream 
have Gaia proper motions contained within the distribution shown
in the figure, and which also lie in the main sequence area of the CMD as above (Fig \ref{fig:cmd_observed}).  A catalog was thus created, and used in the Hectochelle observations that we next describe.

\section{2019 Hectochelle Data}\label{chelle}
In the latter stages of composing this paper, some additional MMT telescope time
became available which was used again with the Hectochelle instrument \citep{saint} , in order 
to obtain high resolution data for abundance analysis. That work will be reported
elsewhere, but we can discuss additional members found in that data for this paper. The observing
program consisted mainly of stars in the denser part of the stream already known to be members, but we also added some previously unobserved stars that satisfied
the CMD and proper motion criteria described above, with a preference for stars
on the giant branch rather than the MSTO because of the higher spectral resolution 
being observed. The observing setup was similar to that for
the \cite{sesar2015} data except that a different wavelength region was used : 6135 to 6320 \AA . The total exposure time on the single 1\degr \ field was 15,700s.
80 stars were targeted (these are also shown in Fig \ref{fig:cmd_observed},
34 of which provided good velocities. Fortuitously, 9 more members were found
in this data, including the brightest member currently known, with $i=14.8$,
which is on the RGB at the brightness level of the HB. All of these members have radial velocities
much more precise than the Hectospec, Binospec, or Deimos data, with the mean 
velocity uncertainty being 0.8 \kmsc.

\section{Using the {\it EM} technique to determine membership}\label{em}

In summary, we have used three different instruments on the MMT telescope to obtain new spectra and thus radial
velocities for targeted stars over a $6\degr \ $ RA extent in the OphStr region. Successful low resolution spectra with
uncertainties of order 15 \kms were obtained for 900 stars using Hectospec and 400 stars using Binospec.
Higher resolution Hectochelle data were successfully obtained for 34 more stars, which have uncertainties
of order 0.8 \kmsc. These radial velocities can of course be used to determine membership, but we can also add in GDR2 proper motions.

The relation between proper motion and RA for radial velocity member stars shown in Fig \ref{fig:oph_pm8} wase used to determine membership, along with positions and radial velocities. These relations were thus incorporated into the statistic used by our ``expectation maximization'' (EM) code, patterned on
that of \cite{walker09}. EM is an iterative algorithm for estimating the mean and variance of a distribution sampled with contamination. The EM method differs from a conventional, sigma-clipping method in two key respects. First, whereas the conventional method answers the question of membership with a yes or no, the EM method yields estimates of model parameters while evaluating membership of individual stars probabilistically. No data points are discarded; rather, likely contaminants receive appropriately small weights. Second, in evaluating membership probability, the EM method explicitly considers the distributions of both members and contaminants, field stars in this
case. Thus, likely contaminants that happen to lie near the center of the member distribution can be identified as such and weighted accordingly. EM therefore allows a full likelihood treatment of the entire data set.

EM can take various observables and select groups from those. It was our task then to 
determine which observables to use. Aside of the radial velocities, we had available the positions, 
proper motions, parallaxes and spectral line indices for the Hectospec data. We show below that the
parallaxes are at the edge of significance, and thus they
did not add any value to membership
determination. Likewise, the 
spectra are fairly low S/N, and while the spectrum formed by averaging the spectra of
members determined with radial velocities and proper motions is clearly different from that of
field stars (the stream members have stronger H$\beta$ and weaker Mgb lines,
in concordance with a metal-poor population), individual spectra are not good enough to use for spectral analysis, and hence do not contribute significantly
to membership determination.
Thus we employed only the radial velocities, the proper motions, and the distance from a curved line on the sky 
(which would be the mean stream position) in the membership search. The location of the stars in the color-magnitude diagram already entered into
the selection for spectroscopic observation and would not add any further information in the analysis, and so was excluded.
For position, we used the departure from a quadratic equation in the coordinates, rather
than distance from a single point as would be the case for a star cluster or dwarf galaxy. 
In doing so, we explicitly lower the probability of membership 
for stars that are farther away from this curve, meant to represent the densest part of the
stream.
For
proper motion, we allowed the mean proper motions to have a dependence on coordinates, as 
discussed in Section \ref{gaia}. In total, we solve for 24 parameters. For members and non-members, we seek a mean radial velocity, mean proper motions, the spread or scale of those values, and their gradients with the two coordinates. 
Following the description of the method in \cite{walker09}, we initialize the probabilities at 0.5 for all stars, and solve the equations for the parameter estimates which are similar to Eq. 13 and 14 of \cite{walker09}. New probability estimates are then calculated, and the procedure is repeated until stability is reached. We typically iterated for 500 steps.
In this analysis, there are five stars whose radial velocity would indicate membership, but whose GDR2 data indicate they are not members, not surprising given the distribution of the field 
stars shown in Fig. \ref{fig:vel}. Similarly, there are four other stars whose radial velocity would indicate membership (though the errors are large), but which have no data in the GDR2, hence they are also not considered members here.

To estimate the uncertainty of the derived parameters, we further performed a Monte Carlo experiment drawing 1000 smples from the input data set, and running the EM calculation again. The samples
were created by creating new samples from the full catalog, with replacement, and using the star's formal uncertainties in radial velocity and proper motion.  The standard deviations of the results for the 24 parameters from the 1000 runs were then calculated. The uncertainty for the velocity dispersion from this
  heterogenous data set is
not trustworthy, nor is the stream dispersion itself, as we discuss below.

\begin{deluxetable*}{lrrrr||rrrr}
\tablecolumns{9}
\tabletypesize{\scriptsize}
\tablewidth{7.0truein}
\setlength{\tabcolsep}{0.1in}
\tablecaption{EM Output\label{em_output} }
\tablehead{\colhead{Parameter} &\colhead{Mean} &\colhead{Disp.} & \colhead{Grad $\xi$\tablenotemark{a}}  & \colhead{Grad $\eta$\tablenotemark{b}}   &\colhead{Mean} &\colhead{Disp.} & \colhead{Grad  $\xi$\tablenotemark{a}}  & \colhead{Grad  $\eta$\tablenotemark{b}}   \\
\hline
\colhead{} &\multicolumn{4}{c}{Members}& \multicolumn{4}{c}{Nonmembers}   \\
}
\startdata
&\kms &\kms &\kms &\kms &\kms &\kms &\kms &\kms \\
& & &arcmin$^{-1}$&arcmin$^{-1}$& & &arcmin$^{-1}$&arcmin$^{-1}$\\
Velocity & 284.4 & 22.3 & -0.019 & 0.16 & -26.4 & 121.4 & -0.067 & -0.166\\
error & 2.0  & 2.5  & 0.021  & 0.15  & 3.02  & 3.86  & 0.025  & 0.093 \\
\hline
& mas~yr$^{-1}$ &mas~yr$^{-1}$ &mas~yr$^{-1}$ &mas~yr$^{-1}$ &mas~yr$^{-1}$ &mas~yr$^{-1}$ &mas~yr$^{-1}$ &mas~yr$^{-1}$  \\
& &  &arcmin$^{-1}$&arcmin$^{-1}$& & &arcmin$^{-1}$&arcmin$^{-1}$ \\
$\mu_{\alpha}$ &-6.63 & 0.99 & -0.015 & -0.015 & -4.02 & 4.63 & -0.0010 & -0.010\\
error &  0.11  &  0.09  & 0.001  &  0.009  &  0.12  &  0.09  &  0.0010  & 0.004 \\
$\mu_{\delta}$ &-4.37 & 0.73 & 0.0010 & 0.001 & -5.16 & 4.46 & -0.0040 & -0.004\\
 error &  0.06  &  0.07  &  0.0007  &  0.007  &  0.10  &  0.08  & 0.0009  & 0.003 \\

\hline
\enddata
\tablenotetext{a}{Gradient in the RA coordinate}
\tablenotetext{b}{Gradient in the Dec. coordinate}
\end{deluxetable*}

Table \ref{em_output} contains the output of the EM technique, while Table \ref{spectra} contains a column listing the membership probabilities where all the necessary data exist (position, radial velocity, proper motions and their uncertainties). The members and nonmembers are shown in separate groups in Table \ref{em_output}, with the mean, scale or dispersion, and gradients in $\xi$ (RA coordinate) and $\eta$ (Dec coordinate) listed. The main distinguishing features between the
member group and nonmember group have been shown in various figures already, the large mean radial velocity, the small dispersion in radial velocity and proper motion, and the dependence of proper motion on coordinates. There are currently 203 stars in our study with membership probability P$_{\rm mem} > 0.9$, and another 15 with a lower threshold of 0.5.

The previous work of \cite{sesar2015} and \cite{sesar2016} had identified 14 and 4 possible members, respectively. We 
confirm all of these as members except for ``cand26'' from \cite{sesar2016}, which is excluded because its
proper motion is out of variance with the other members, in addition to its radial velocity as noted in \cite{sesar2016}.

\begin{figure*}
  \centering
  \begin{tabular}{@{}ll@{}}
    \includegraphics[width=4.0in, trim=0in 1in 0.5in 1.1in,clip]{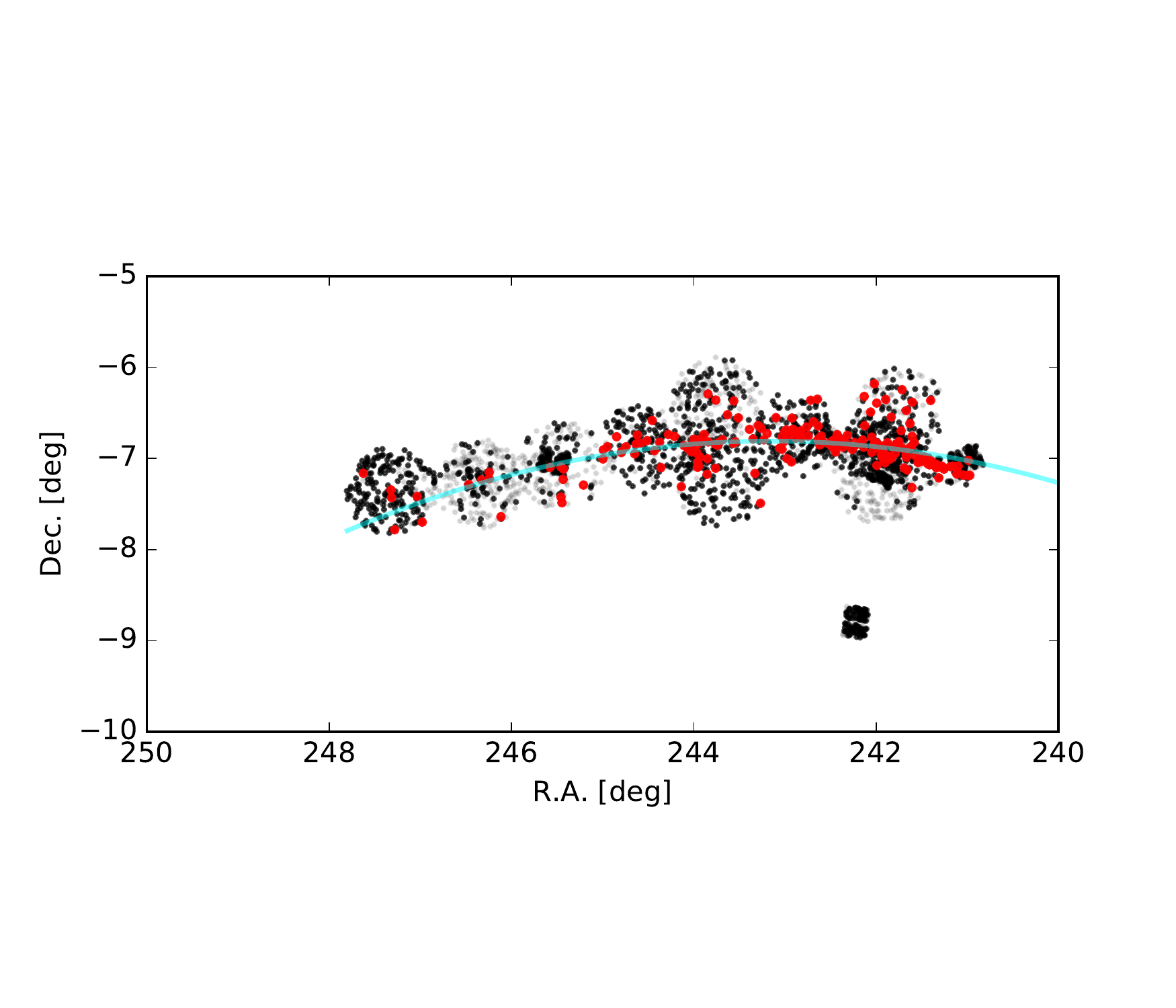}&\includegraphics[width=4.0in,trim=0.8in 0in 0in 0in,clip]{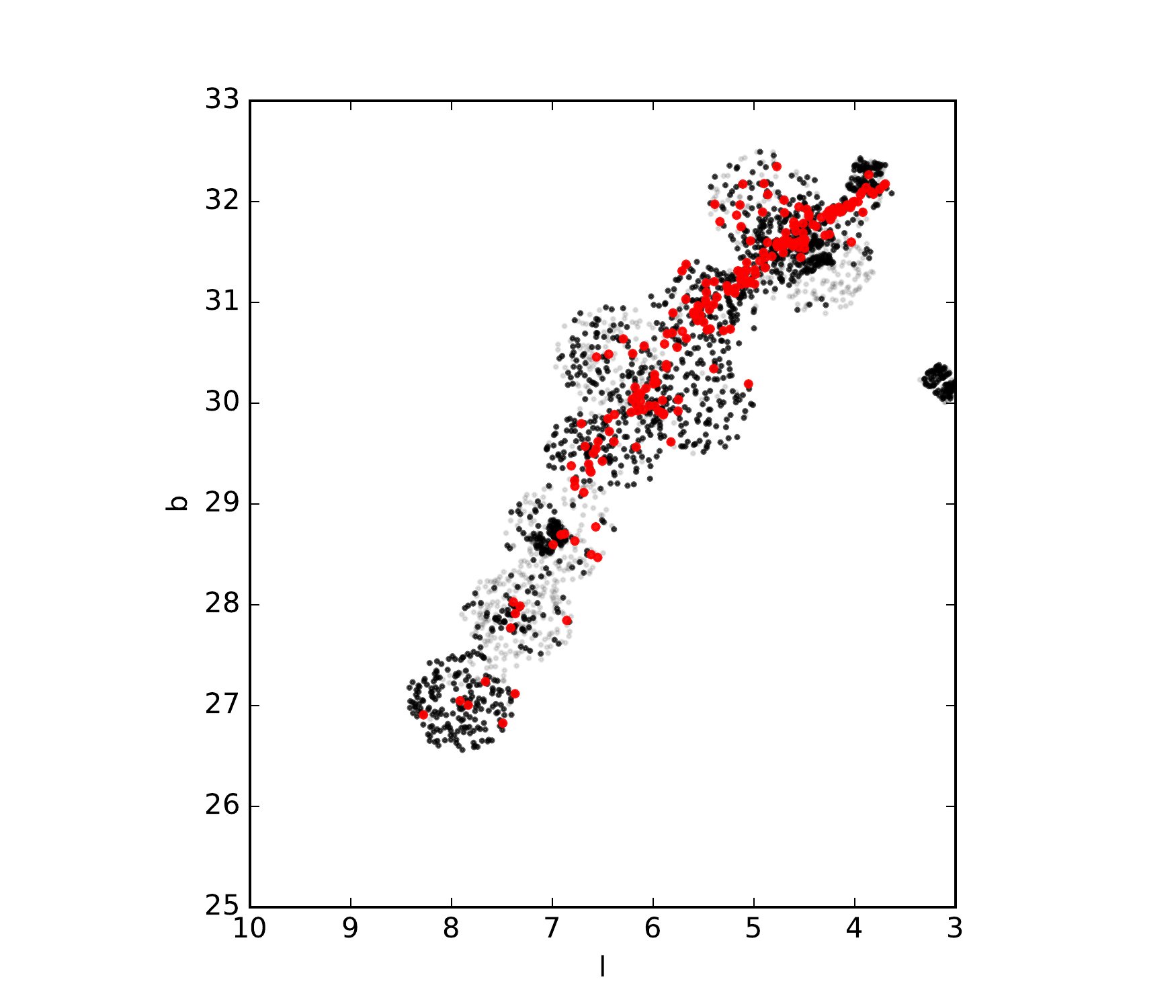}
  \end{tabular}
  \caption{\textit{Left:} Position on the sky of all spectroscopic observations. The light grey circles are 
observations that did not result in accurate radial velocities, due to signal-to-noise problems. Dark 
circles are successful observations. Red circles show the members found by the EM technique
(where  P$_{\rm mem} > 0.9$), 
using radial velocities and proper motions. The solid line is the calculated mean distribution in the 
form of a quadratic equation, for
use in EM (it is not meant to represent a possible orbit). Note the areas where coverage with
good observations was poor, but also note the scatter away from the mean distribution for members.    
    \textit{Right:} Same, in galactic coordinates. We omit the line.
    \label{fig:stream_pos}}
\end{figure*}

Fig. \ref{fig:stream_pos} shows the distribution of observations on the sky. Light gray 
circles are those observations for which we could not obtain useful radial velocities, due to low
signal-to-noise. Dark circles are used for the successful observations, while red circles
shows the members with P$_{\rm mem} > 0.9$ found using the EM technique (90\% of those
stars have probabilities greater than 0.95). 

We find the following general results. The stream extends over
the entire region we observed spectroscopically, from RA=241 to 247\degr \ (or $l=3.8 $ to $ 8\degr \ $ and
$b=27$ to $32\fdg2\ $).  While there
is a tight stream as seen before, there are also some high probability stream members 
significantly away from
the main stream. There is no significant detection of a radial velocity gradient (thus perhaps
requiring better data to see such a thing. Finally,  there is a significant gradient in the RA component
of the proper motion with RA itself, as we noted in Fig. \ref{fig:oph_pm8}. The stars away from
the tight stream appear to have the same spectral characteristics as those in the main stream
(and different from the non-member stars), though
higher signal-to-noise spectra would be needed for quantitative analysis.

One of the fields observed with Binospec was a control field, approximately $2\degr \ $ south
of the densest part of the stream. Only one star in that field has P$_{\rm mem} = 0.5$ (and thus not
shown in \ref{fig:stream_pos}, and is thus likely not a member of the OphStr. Thus the stream does not
extend to the very large angle of $2\degr \ $.

Fig. \ref{fig:oph_pm8} shows the proper motion relations when we use EM membership criteria for the Hectospec and previous data (black points only). The outliers are removed. This then allowed us to determine where best to place new observational fields, with
the smaller field of Binospec (and later with more Hectochelle observations).
Fig \ref{fig:oph_pm8} also shows the relation when we use EM membership criteria, now including the Binospec data (the blue points).

 The velocity dispersion derived for the stream with this heterogeneous data set is $22\pm2.5$ \kmsc . However, the use of low precision data in determining small velocity dispersions is known to result in systematically higher dispersions than true \citep{kop}. To avoid this issue, we rederived the velocity dispersion using only the Hectochelle data for 18 members, which is homogeneous and for which each star has a velocity
uncertainly of $< 2.6$ \kmsc . 
Our derived velocity dispersion using those stars, which are in the densest part of the stream
and in the area from $4.2 < l < 5.0\degr \ $ is $1.6 \pm 0.3 $\kmsc. In Table \ref{spectra},
these are the stars that have P$_{\rm mem} > 0.9$ and which 
have either ``F'' or ``E'' in the first column. This value is
larger than reported in \cite{sesar2015}, even though we reuse 9 of the stars in
that paper, along with our 9 new members with similarly precise radial velocities. 
Indeed, the reported \cite{sesar2015} value of
$0.4 ^{+0.5}_{-0.4}$ \kms was in fact undetermined.

For orbit analysis,
we now have stellar membership probabilities, and can use those values to refine an estimate of the cluster's length and orbit, which we take up in Section \ref{orbit}.

\subsection{A Gaia parallax?}

We then explored what the GDR2 parallax catalog could tell us about the distance.
\cite{leung} and \cite{schonrich} have suggested that the Gaia parallaxes should have $+0.05$ mas added to GDR2 tabulated values, which
we will incorporate.

Recall that from modelling the photometric data for 14 member stars, \cite{sesar2015} had derived a distance modulus of 14.58 and a gradient in the modulus as well, such that 
the stream at $l=4$ (perhaps the leading edge of the stream) has a distance of 9 kpc, while
2\degr \ away at $l=6$, the distance was derived to be 7.5 kpc. This was confirmed from the CMD
of the new member stars in Section \ref{cmd_distance}. From our final members
list, we downselected to include stars where the formal parallax errors were less than
0.5 mas and which had membership probabilities greater than 0.9. 
This leaves us with 118 stars spread over 4\degr \ in longitude \citep[i.e., twice as
long as][]{sesar2015}. The weighted mean GDR2 parallax of this list is $0.12\pm0.01$ mas, giving a distance of 8.3 kpc, while that of the 1308 non-members
is $0.44 \pm 0.004$ (2.2 kpc; there are a number of bright stars in this group that are nearby). 

What about the distance gradient along the stream? We can 
form an average parallax for stars grouped at $l=4$ and 6\degr \ as \cite{sesar2015} considered. The 4\degr \
group has 22 stars, with a weighted mean parallax of 0.15, and mean error of 0.07. while
the 6\degr \ group with 29 stars has a parallax of 0.12 and error of 0.07. Thus, neither
is significant though the gradient is in the same sense as the photometry indicates.  
For the more extended stars found here, at $l>$7\degr, we have
only 7 stars, whose weighted mean parallax is 0.08, and error of 0.05. The distance change again
agrees with the photometry, though the significance is small. 

The two brightest member stars (with i $< 15.3$), one RGB star and one BHB star, 
are located in the dense part of the stream, and $3\fdg4\ $ to the east in the 
low density area, respectively.
These two stars individually have significant detections
of parallaxes (at 3.5 and 3.0$\sigma$ respectively). The average of these is 0.20+/-0.03 mas for a distance of 5.0 kpc (the two stars have similar parallaxes). It appears that the mean GDR2 parallax of radial velocity members is in agreement with the distance from the CMD, but does not add any significance to the measurement.

\section{a new map using only Gaia and PS1 data}\label{newmap}


\begin{figure}[ht]
 \includegraphics[width=4.0in]{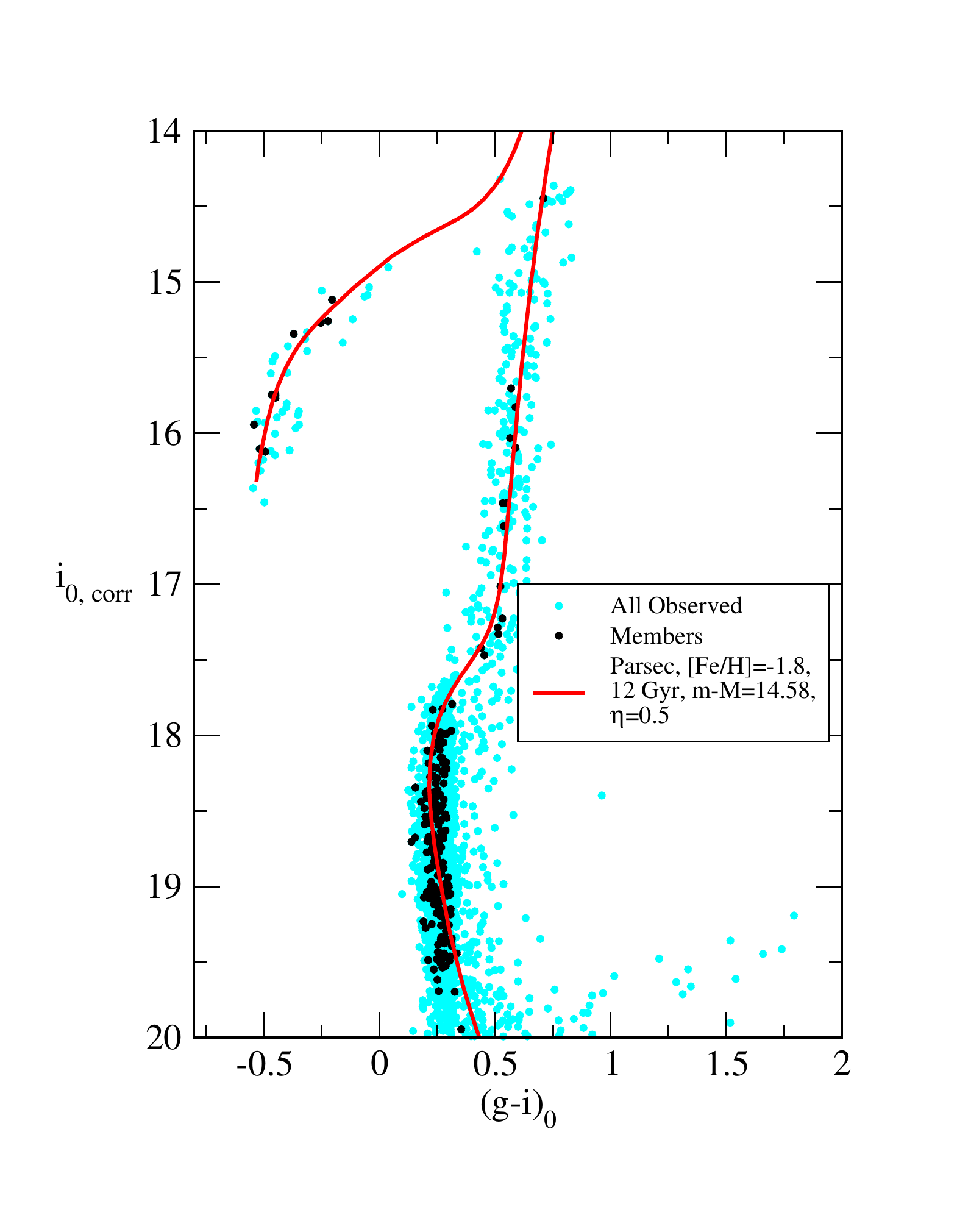}
\caption{CMD of all stars observed from all data sets including the Binospec data (which are not in Fig. \ref{fig:cmd_observed}),  shown as cyan points. Stars with a high probability being members are shown in black. Photometry and isochrone information is the same as for Fig \ref{fig:cmd_observed}. 
\label{fig:cmd_observed_all}}
\end{figure}

Fig \ref{fig:cmd_observed_all} is a simple version of the CMD, with the all input samples shown with the same color, and the members as calculated from Section \ref{em} shown in black. This figure demonstrates that the input sample did not bias the selection of members in the CMD plane, and that the PARSEC isochrone selected in \cite{sesar2015},
along with the distance modulus and gradient is still valid for the members.

Fig \ref{fig:stream_pos} highlights two parameters of the stream. First the stream is longer than
initially realized in the \cite{bernard} paper. It extends more than 6\fdg5\ in RA rather
than 2\fdg5\ . Second, for the western part, there is a tight stream, which is what  \cite{bernard} found, but there are also members scattered about that tight area in a more diffuse halo. 
The extended eastern part of the stream is also diffuse. 
Because the observing conditions for the observations varied so much, and the fields were not all
contiguous, it was not possible to make a satisfactory diagram of the 2D distribution using
just the member stars as derived in Section \ref{em}. For instance, does the 
sprinkling of member stars to the north of the dense stream at RA$=242\degr \ ( l=5\degr \ $) have a symmetrical distribution to the south? The radial velocity data is not complete enough to answer that. We can 
alleviate the problems produced by the spatial incompleteness of the radial velocity data by making a revised
map using the positions and photometry of likely members from a stellar catalog such as PS1 combined
with proper motions from GDR2. While the stream's proper motions do not readily distinguish it from the other stars in the field (see Fig \ref{fig:pm_both}) as well as the radial velocity measurements do, the mean motion does provide an additional selection criterion. We used the refined CMD locus from the known members and the GDR2 catalog and were able to produce an image that does better reflect the large-scale distribution of stars than revealed
in the radial velocity measurements.

\begin{figure*}[ht!]
\includegraphics[width=7.25in]{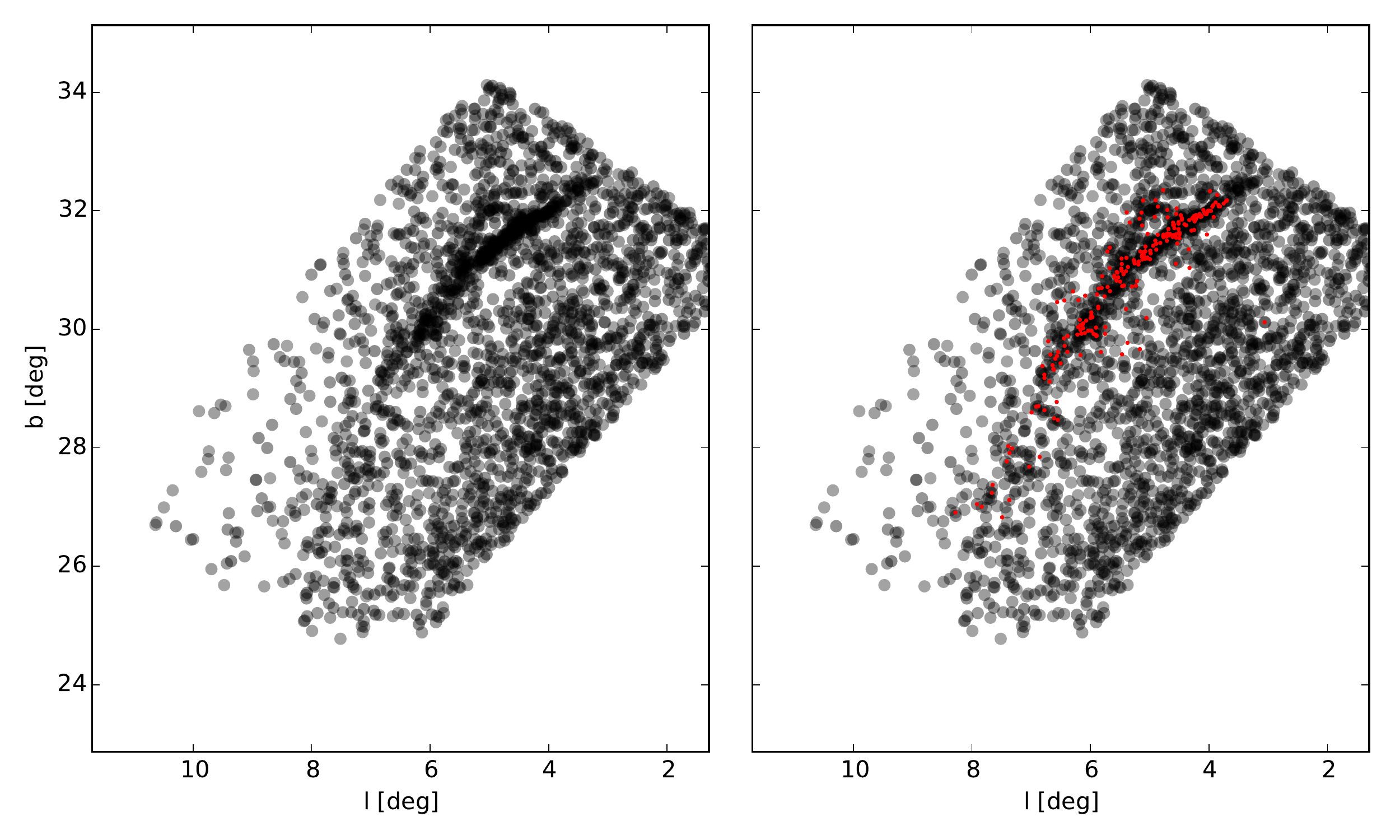} 
\caption{Using the measured proper motion and distance modulus gradient for the stream,
we selected HB, MSTO and lower RGB stars from the full GDR2 catalog for display. Probabilities of stream membership
were based on deviation from the isochrone used above, and also from the relation of
proper motion and position determined above. No radial velocities were used. On the right the stars observed spectroscopically
and which have stream membership probabilities of greater than 50\% are shown in red, revealing there to be good correspondence, even for stars away from 
the main stream.
\label{fig:gaia_dist}}
\end{figure*}

In detail, we assigned membership probabilities from the PS1 and GDR2 catalogs. For the
CMD filtering, the best fitting
isochrone remains the one from PARSEC, with [Fe/H]$=-1.8$, age=12\,Gyr, $\eta=0.5$, m$-$M$=14.58$, and 
with the \cite{sesar2015} distance modulus gradient, m$-$M$=14.58 - 0.2({\it l} -5$). For a 
star with color and magnitude, ($g-i$)$_0$ and i$_{0,corr}$, the difference between the observed color and that
of the isochrone at the star's i-band magnitude was calculated, and turned into a probability via abs(($g-i$)$_{0}-$($g-i$)$_{isochrone}$)/$\sigma_{rel}$, where $\sigma_{rel}$ 
is the spread in the observed catalog about the isochrone, at magnitude i$_{0,corr}$. 
We 
used stars in the region of the HB, the main sequence brighter than i$_{0,corr}=20.0$ (i$=20.4$) , 
and the giant branch
up to about 1 magnitude below the level of the HB, which was about i$_{0,corr}=15.5$ (or observed 
i$=15.8$).
Proper motions
were used in a similar fashion, this time using the distance to the mean relation of 
proper motion with coordinates as defined by the radial velocity members (Fig. \ref{fig:oph_pm8}) divided
by the spread in that relation. These quantities were summed quadratically along with the CMD probability 
for each star,
and turned into a transparency value for graphical purposes. Fig. \ref{fig:gaia_dist} shows
the result, in galactic coordinates. The CMD filtering alone provides most of the contrast
in this diagram, but the addition of the proper motion filtering eliminates much of the remnant dust
dependency of the CMD filter. Adding in the giant stars to the MS and BHB list did not 
add much contrast to this figure; only ten percent of the stars shown are giant
candidates.  Removing stars with accurate parallaxes indicating non-membership also
did not achieve much; there were only 50 such stars. In total, there are about 2400 stars remaining after the selection criteria were applied.

Fig. \ref{fig:gaia_dist} confirms the map made using the additional parameter of the radial velocities: a nearly-linear, tight stream extending from $l=$3\fdg2\ to 5\fdg8\ (380 pc), and a longer more diffuse 
tail with more curvature (in these projected coordinates) extending as far as $l=8$\degr \ for a total length of 7.5\degr , or 1.1 kpc in projection.  The leading
edge of the stream in the radial velocity membership is $l=$3\fdg8\ , but the proper-motion/CMD map alone shows the distinct edge to actually be at RA$=239.8$, DEC$=-7.24$ ($l=3\fdg2 \ , b=32\fdg4 \ $), which is $0\fdg5 \ $ farther from the
edge of the spectroscopic coverage.
If the leading edge also has a diffuse component, it is possible that the stream extends beyond our catalog's footprint, but we leave mapping the full extent of the stream to future work. 
Using the same distance modulus gradient as found 
in \cite{sesar2015}, the deprojected length of the currently-detected stream is about 3 kpc. 

Finally,  Fig. \ref{fig:gaia_dist} confirms the presence, noticed in the radial velocity work, of members in
 a curious arc-like structure that arches in latitude away from the main stream at $l=4$,
to a separation of nearly a degree, and then apparently rejoins the main stream at
$l=6$.  Recall that for the EM analysis above, stars farther from the main stream are
explicitly judged to have lower membership probabilities, so the number of EM member stars
in this spur may be an underestimate. A topic for future investigation would be
whether this sub-stream is similar to the spurs on GD-1 and Jhelum found and discussed 
in \cite{price18}, \cite{bonaca1} and \cite{bonaca2}.

\section{Orbit fitting}\label{orbit}

\begin{figure}
\begin{center}
\includegraphics[width=3.0in]{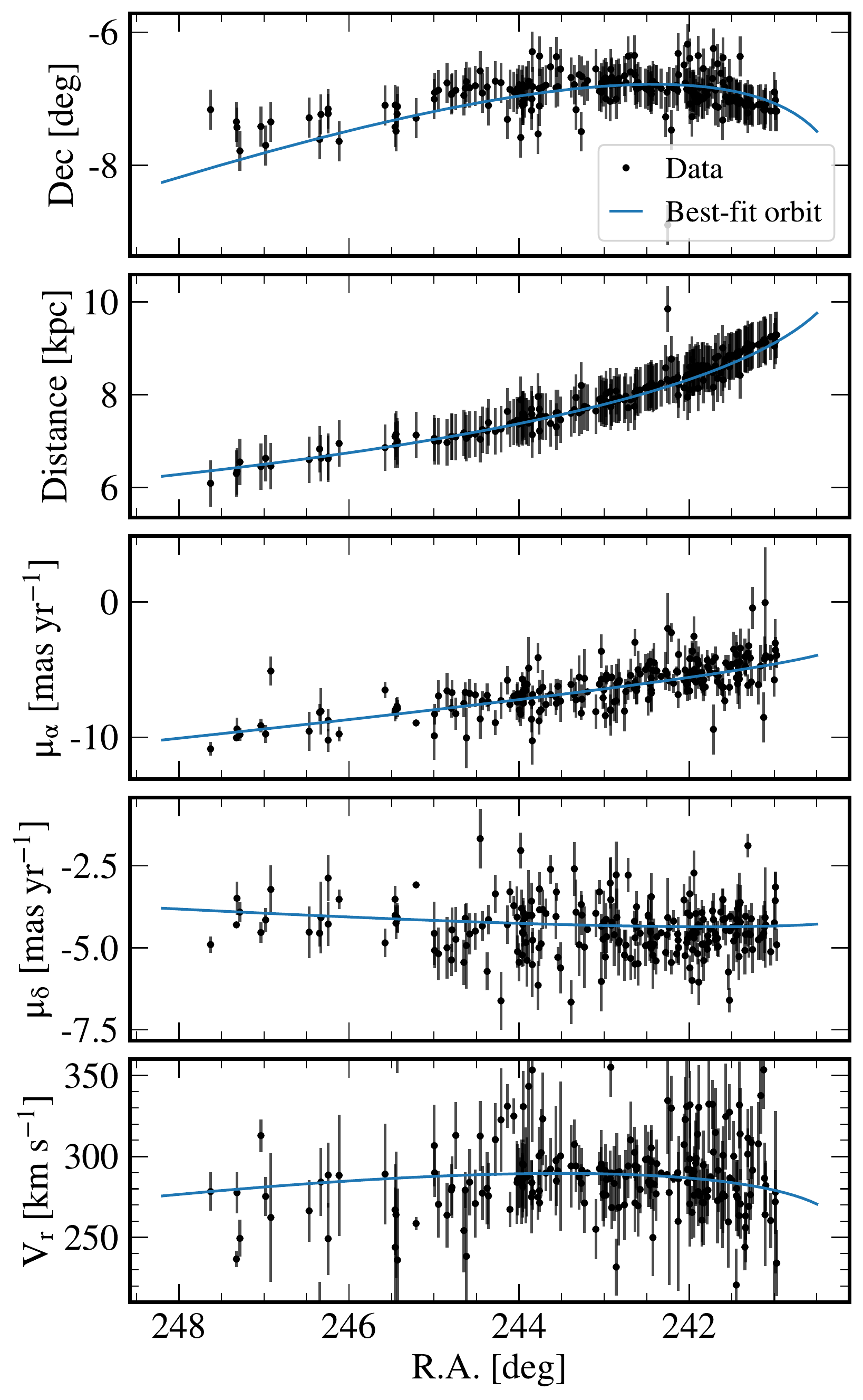}
\end{center}
\caption{Phase-space distribution of likely Ophiuchus member stars (black points) and the best-fit orbit in a smooth and static Milky Way potential (blue lines).
}
\label{fig:orbitfit}
\end{figure}

\begin{figure}
\begin{center}
\includegraphics[width=3.0in]{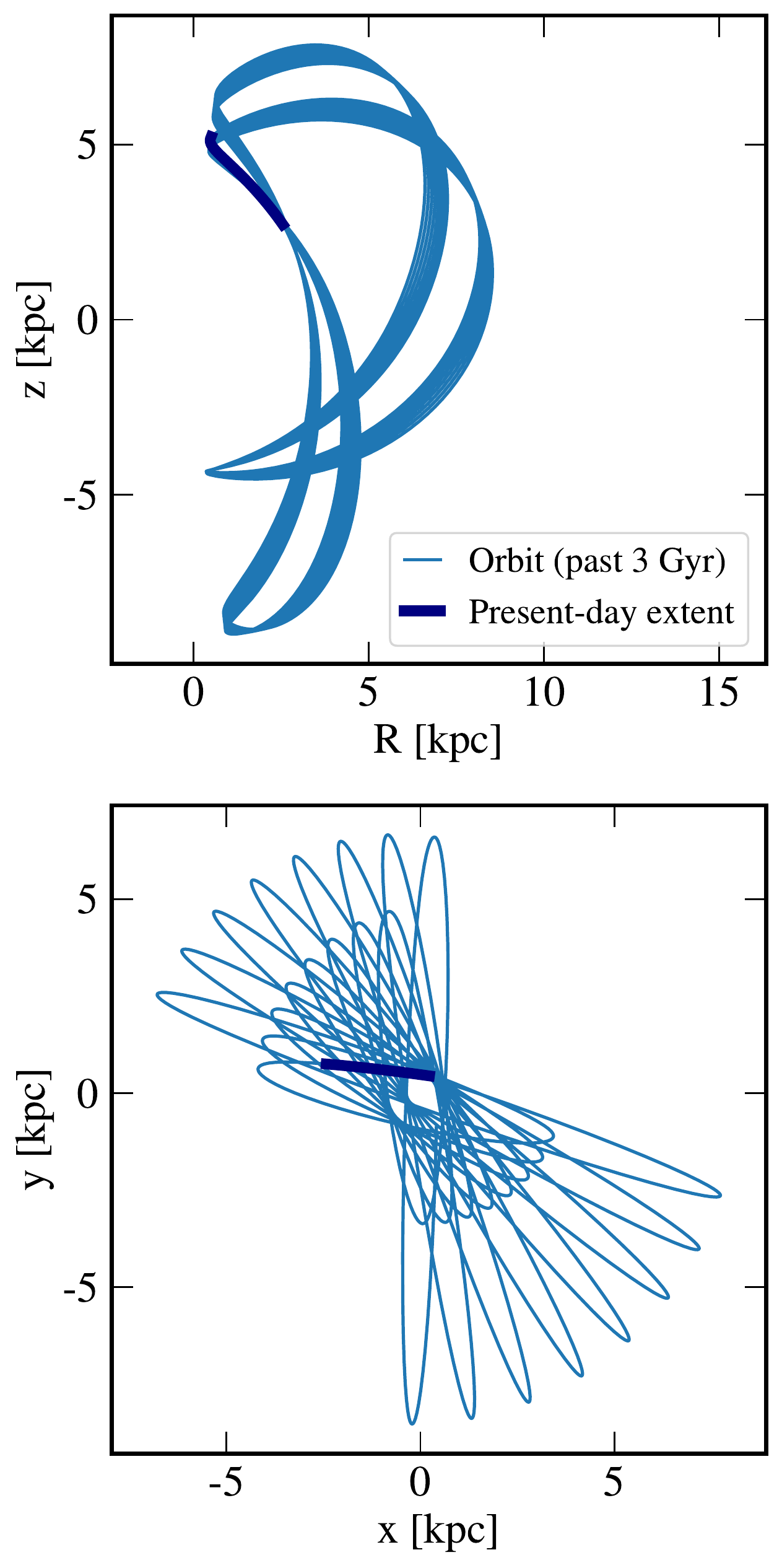}
\end{center}
\caption{The best-fit orbit to the Ophiuchus stream in the Galactocentric coordinates perpendicular to the disk plane (top), and in the disk plane (bottom). The Sun is at z=0, R=8 kpc in the top
plot.
Current extent of the stream is shown in dark blue, while the past 3\,Gyr are in light blue.
}
\label{fig:orbit}
\end{figure}

In this section we revise the orbital properties of OphStr given the newly detected extension of the stream.
Figure~\ref{fig:orbitfit} summarizes the 6D phase-space distribution of likely Ophiuchus members.
From top to bottom, we show Dec, distance, RA and Dec components of the proper motion vector, and radial velocity as a function of RA, with the errorbars representing the observational uncertainties.
The proper motion uncertainties are delivered by GDR2, while we adopted the stream width of $0.3^\circ$ as the uncertainty in sky positions, and assumed $0.5\,\rm kpc$ uncertainty in distances, and $10\,\rm km\,s^{-1}$ for radial velocities.
We use these data to search for plausible orbits of the OphStr progenitor.

In our modeling, we used a static Milky Way potential that has a $5.5\times10^{10}\rm\, M_\odot$ \citet{mn1975} disk (scale height 28\,pc, scale length 3\,kpc), a $4\times10^9\rm\, M_\odot$ \citet{hernquist1990} bulge (scale radius 1\,kpc), and a $7\times10^{11}\rm\, M_\odot$ \citet{navarro1997} halo (scale radius 15.62\,kpc, $z$-axis flattening 0.95), as implemented in the \package{gala} package \citep{gala}.
To convert between the physical and observed (heliocentric) quantities, we adopted a right-handed coordinate system with the origin at the Galactic center in which the Sun is located at $(X,Y,Z)=(-8.122,0,0.0208)\,\rm kpc$ and is moving at velocity $(V_X, V_Y, V_Z)=(12.9, 245.6, 7.78)\,\rm km\,s^{-1}$ (available in \package{astropy} version 4.0).
We searched for a best-fitting orbit in this gravitational potential by minimizing the deviations between the orbit and individual stars using the \package{scipy} implementation of the BFGS algorithm.
We evaluated trial orbits at RA positions of likely Ophiuchus members, taking into account the observational uncertainties.
The best-fit orbit is shown with blue lines in Figure~\ref{fig:orbitfit}.

The orbit which was evolved in a simple, static Milky Way potential captures the global phase-space distribution of the OphStr stream, including the newly discovered extension at $\rm RA\gtrsim244^\circ$.
The best-fit orbit reproduces well the observed gradients in distance, proper motions and and radial velocity.
This orbit is somewhat misaligned from the observed stream track, as expected of streams in general \citep[e.g.,][]{sanders2013}.

The past 3\,Gyr of Ophiuchus' orbit are shown in Figure~\ref{fig:orbit} in light blue, while the current extent of the stream is emphasized in dark blue.
 The stream is just past the pericenter, which occurred at $\approx3.8$\,kpc, in agreement with previous orbit determinations using the densest part of the stream \citep{sesar2015,lane2019}.
However, we inferred a smaller apocenter of 13.5\,kpc than previously derived 16.8\,kpc, which directly translates to a shorter period of 172\,Myr (compared to 240\,Myr).
Generally, we expect longer streams to be more informative \citep{bh2018}, and therefore our orbit which fits a longer segment of the stream to be more accurate.
In addition, this is the first self-consistent orbit fit using the accurate Gaia proper motions, which might contribute to the change in the derived orbit.
Fits to the latest Ophiuchus data place its progenitor on a somewhat less eccentric orbit than previously estimated, so dynamical models aiming to reproduce the observed stream should be revised to use this new orbit.

\clearpage
\section{discussion and conclusions}
Our spectroscopic work has increased the number of probable members (P$_{\rm mem} > 0.9$) of the Ophiuchus stream from the 18 reported previously to about
200, which lie along an arc 7.5\degr \ long in projection, or about 3 kpc deprojected. 
In finding these new members and candidates however, we have not significantly increased
the estimated luminosity of the progenitor cluster found by \cite{bernard} of M$_{\rm V}=-3.0$,
or the estimated mass of $2\times 10^4$ M$_\sun$ from \cite{sesar2015}. Using our new method for finding probable members
from the GDR2 and PS1 data (Sect. \ref{newmap}), and a correction provided
by the luminosity function for the Parsec isochrone used
 in Sect. \ref{hecto} we derived fewer total stars brighter than i=20 than \cite{sesar2015} (thus our
calculated surface brightness is lower), resulting in a new calculated luminosity of the entire stream 
of M$_V=-3.1$, and a total mass of $8700 \pm 400$ M$_\sun$, where the uncertainty
comes from counting statistics only, and is likely an underestimate. 

 While the new low resolution spectra have allowed us to explore the distribution
of members and refine orbit models, few of the derived radial velocities are as precise as
one would need for mass estimates. In particular, the low resolution data appears to allow
the velocity variation with longitude outlined here as well as those in \citet[][their figure 2]{adrian} or \citet[][their figure 3]{lane2019}. 
To independently constrain the mass of the OphStr progenitor, we only use the high-resolution Hectochelle data and obtain a new velocity dispersion value of $1.6\pm0.4$ \kms.
Empirically, we expect the velocity dispersion in the cluster to be similar or lower than in its tidal tails \citep[for example, the dispersion in the Palomar~5 globular cluster is 1.2\,\kms and 2.1\,\kms in its tails,][]{odenkirchen2009, kuzma2015}.
There are six clusters in the \citet{harris} catalog with velocity dispersions equal to or less than 1.6\,\kms, however, these are all somewhat more luminous than our value for OphStr.
However, dynamical heating of tidal tails may resolve this inconsistency between the inferred mass and velocity dispersion.
Specifically, \citet{lane2019} simulated globular clusters with a range of masses and sizes on OphStr orbits, and their more diffuse progenitors of $\approx8700\,M_\odot$ achieve a velocity dispersion of $>1\,\kms$ after three complete orbital periods.
Similar simulations that include our new constraints on the OphStr extent and radial velocity dispersion will further refine our initial estimate of its progenitor's mass.



 The stream orbit as refined by this new data is smaller than previously determined, with a period of 172 Myr and an apocenter of 13.5 kpc.
This would indicate that the galactic bar should have an even larger influence on the stream dynamical evolution than previously thought. 

Similar to previous papers on OphStr, we did not find a progenitor star cluster for OphStr, unlike the case for the Palomar 5 stream \citep{price19}. 
 Interestingly, we did find evidence for vertical sub-structure in the form of a spur.
\cite{bonaca2} showed that spurs can be produced if a stream encounters a massive object, such as a dark-matter subhalo or a globular cluster.
The abundance of dark-matter subhalos is expected to be severely depressed in the inner Galaxy due to disk disruption \citep[e.g.,][]{donghia2010, errani2017, gk2017}, but most Galactic globular clusters orbit in the same volume as OphStr \citep[e.g.,][]{baumgardt2019}.
Precise kinematic maps of the affected region in the stream can constrain the impactor's orbit \citep{bonaca2020}, motivating additional high-resolution spectroscopy of OphStr to uncover its complex dynamical history.


\software{
    \package{Astropy} \citep{astropy, astropy2018},
    \package{gala} \citep{gala},
    \package{matplotlib} \citep{mpl},
    \package{numpy} \citep{numpy},
    \package{scipy} \citep{scipy}
}

\acknowledgments
We thank Martin Paegert, Sean Moran, and the MMT remote observers for help in obtaining the telescopic and archival data. We also thank Christian Johnson, Gyuchul Myeong, John Raymond, Joel Pfeffer, and Julio Navarro for illuminating
conversations. NSF Grant AST-1812461 supported work on this project.

\clearpage



\begin{thebibliography}{71}
\expandafter\ifx\csname natexlab\endcsname\relax\def\natexlab#1{#1}\fi

\bibitem[{{Astropy Collaboration} {et~al.}(2013){Astropy Collaboration},
  {Robitaille}, {Tollerud}, {Greenfield}, {Droettboom}, {Bray}, {Aldcroft},
  {Davis}, {Ginsburg}, {Price-Whelan}, {Kerzendorf}, {Conley}, {Crighton},
  {Barbary}, {Muna}, {Ferguson}, {Grollier}, {Parikh}, {Nair}, {Unther},
  {Deil}, {Woillez}, {Conseil}, {Kramer}, {Turner}, {Singer}, {Fox}, {Weaver},
  {Zabalza}, {Edwards}, {Azalee Bostroem}, {Burke}, {Casey}, {Crawford},
  {Dencheva}, {Ely}, {Jenness}, {Labrie}, {Lim}, {Pierfederici}, {Pontzen},
  {Ptak}, {Refsdal}, {Servillat}, \& {Streicher}}]{astropy}
{Astropy Collaboration}, {Robitaille}, T.~P., {Tollerud}, E.~J., {et~al.} 2013,
  \aap, 558, A33

\bibitem[{{Astropy Collaboration} {et~al.}(2018){Astropy Collaboration},
  {Price-Whelan}, {Sip{\H{o}}cz}, {G{\"u}nther}, {Lim}, {Crawford}, {Conseil},
  {Shupe}, {Craig}, {Dencheva}, {Ginsburg}, {Vand erPlas}, {Bradley},
  {P{\'e}rez-Su{\'a}rez}, {de Val-Borro}, {Aldcroft}, {Cruz}, {Robitaille},
  {Tollerud}, {Ardelean}, {Babej}, {Bach}, {Bachetti}, {Bakanov}, {Bamford},
  {Barentsen}, {Barmby}, {Baumbach}, {Berry}, {Biscani}, {Boquien}, {Bostroem},
  {Bouma}, {Brammer}, {Bray}, {Breytenbach}, {Buddelmeijer}, {Burke},
  {Calderone}, {Cano Rodr{\'\i}guez}, {Cara}, {Cardoso}, {Cheedella}, {Copin},
  {Corrales}, {Crichton}, {D'Avella}, {Deil}, {Depagne}, {Dietrich}, {Donath},
  {Droettboom}, {Earl}, {Erben}, {Fabbro}, {Ferreira}, {Finethy}, {Fox},
  {Garrison}, {Gibbons}, {Goldstein}, {Gommers}, {Greco}, {Greenfield},
  {Groener}, {Grollier}, {Hagen}, {Hirst}, {Homeier}, {Horton}, {Hosseinzadeh},
  {Hu}, {Hunkeler}, {Ivezi{\'c}}, {Jain}, {Jenness}, {Kanarek}, {Kendrew},
  {Kern}, {Kerzendorf}, {Khvalko}, {King}, {Kirkby}, {Kulkarni}, {Kumar},
  {Lee}, {Lenz}, {Littlefair}, {Ma}, {Macleod}, {Mastropietro}, {McCully},
  {Montagnac}, {Morris}, {Mueller}, {Mumford}, {Muna}, {Murphy}, {Nelson},
  {Nguyen}, {Ninan}, {N{\"o}the}, {Ogaz}, {Oh}, {Parejko}, {Parley}, {Pascual},
  {Patil}, {Patil}, {Plunkett}, {Prochaska}, {Rastogi}, {Reddy Janga},
  {Sabater}, {Sakurikar}, {Seifert}, {Sherbert}, {Sherwood-Taylor}, {Shih},
  {Sick}, {Silbiger}, {Singanamalla}, {Singer}, {Sladen}, {Sooley},
  {Sornarajah}, {Streicher}, {Teuben}, {Thomas}, {Tremblay}, {Turner},
  {Terr{\'o}n}, {van Kerkwijk}, {de la Vega}, {Watkins}, {Weaver}, {Whitmore},
  {Woillez}, {Zabalza}, \& {Astropy Contributors}}]{astropy2018}
{Astropy Collaboration}, {Price-Whelan}, A.~M., {Sip{\H{o}}cz}, B.~M., {et~al.}
  2018, \aj, 156, 123


\bibitem[Balbinot \& Gieles(2018)]{balbinot} Balbinot, E., \& Gieles, M.\ 2018, \mnras, 474, 2479 

\bibitem[{{Baumgardt} {et~al.}(2019){Baumgardt}, {Hilker}, {Sollima}, \& {Bellini}}]{baumgardt2019} {Baumgardt}, H., {Hilker}, M., {Sollima}, A., \& {Bellini}, A. 2019, \mnras, 482, 5138
  
\bibitem[Bernard et al.(2014)]{bernard} Bernard, E.~J., Ferguson, A.~M.~N., Schlafly, E.~F., et al.\ 2014, \mnras, 443, L84

\bibitem[{{Bonaca} \& {Hogg}(2018)}]{bh2018}{Bonaca}, A., \& {Hogg}, D.~W. 2018, \apj, 867, 101
\bibitem[Bonaca et al.(2019a)]{bonaca1} Bonaca, A., Conroy, C., Price-Whelan, A.~M., et al.\ 2019a, \apjl, 881, L37
\bibitem[Bonaca et al.(2019b)]{bonaca2} Bonaca, A., Hogg, D.~W., Price-Whelan, A.~M., et al.\ 2019b, \apj, 880, 38

\bibitem[{{Bonaca} {et~al.}(2020){Bonaca}, {Conroy}, {Hogg}, {Cargile},
  {Caldwell}, {Naidu}, {Price-Whelan}, {Speagle}, \& {Johnson}}]{bonaca2020}
{Bonaca}, A., {Conroy}, C., {Hogg}, D.~W., {et~al.} 2020, \apjl, 892, L37

\bibitem[Bressan et al.(2012)]{bressan} Bressan, A., Marigo, P., Girardi, L., et al.\ 2012, \mnras, 427, 127 


\bibitem[Caldwell et al.(2009)]{caldwell} Caldwell, N., Harding, P., Morrison, H., et al.\ 2009, \aj, 137, 94

  \bibitem[{{D'Onghia} {et~al.}(2010){D'Onghia}, {Springel}, {Hernquist}, \&
  {Keres}}]{donghia2010}
{D'Onghia}, E., {Springel}, V., {Hernquist}, L., \& {Keres}, D. 2010, \apj,
  709, 1138

\bibitem[{{Errani} {et~al.}(2017){Errani}, {Pe{\~n}arrubia}, {Laporte}, \&
  {G{\'o}mez}}]{errani2017}
{Errani}, R., {Pe{\~n}arrubia}, J., {Laporte}, C.~F.~P., \& {G{\'o}mez}, F.~A.
  2017, \mnras, 465, L59

  
\bibitem[Fabricant et al.(2005)]{fab} Fabricant, D., et 
al.\ 2005, \pasp, 117, 1411 
\bibitem[Fabricant et al.(2019)]{fab2} Fabricant, D., Fata, R., Epps, H., et al.\ 2019, \pasp, 131, 075004


\bibitem[Gaia Collaboration et al.(2018)]{gaia} Gaia Collaboration, Brown, A.~G.~A., Vallenari, A., et al.\ 2018, \aap, 616, A1 

\bibitem[{{Garrison-Kimmel} {et~al.}(2017){Garrison-Kimmel}, {Wetzel},
  {Bullock}, {Hopkins}, {Boylan-Kolchin}, {Faucher-Gigu{\`e}re}, {Kere{\v s}},
  {Quataert}, {Sanderson}, {Graus}, \& {Kelley}}]{gk2017}
{Garrison-Kimmel}, S., {Wetzel}, A., {Bullock}, J.~S., {et~al.} 2017, \mnras,
  471, 1709


\bibitem[Grillmair, \& Carlin(2016)]{grill} Grillmair, C.~J., \& Carlin, J.~L.\ 2016, Tidal Streams in the Local Group and Beyond, 87


\bibitem[Haffner et al.(2003)]{haffner} Haffner, L.~M., Reynolds, R.~J., Tufte, S.~L., et al.\ 2003, \apjs, 149, 405 
\bibitem[Harris(2010)]{harris} Harris, W.~E.\ 2010, arXiv e-prints, arXiv:1012.3224
\bibitem[Hattori et al.(2016)]{hattori} Hattori, K., Erkal, D., \& Sanders, J.~L.\ 2016, \mnras, 460, 497
\bibitem[{{Hernquist}(1990)}]{hernquist1990}
{Hernquist}, L. 1990, \apj, 356, 359
\bibitem[{{Hunter}(2007)}]{mpl}
{Hunter}, J.~D. 2007, Computing in Science and Engineering, 9, 90

\bibitem[{Jones {et~al.}(2001--)Jones, Oliphant, Peterson, {et~al.}}]{scipy}
Jones, E., Oliphant, T., Peterson, P., {et~al.} 2001--, {SciPy}: Open source
  scientific tools for {Python}, , .
\newblock \url{http://www.scipy.org/}


\bibitem[Kaiser et al.(2010)]{kaiser} Kaiser, N., Burgett, W., Chambers, K., et al.\ 2010, \procspie, 7733, 77330E 
\bibitem[Koposov et al.(2011)]{kop} Koposov, S.~E., Gilmore, G., Walker, M.~G., et al.\ 2011, \apj, 736, 146
\bibitem[Kurtz \& Mink(1998)]{kurtz} Kurtz, M.~J., \& Mink, D.~J.\ 1998, \pasp, 110, 934

\bibitem[{{Kuzma} {et~al.}(2015){Kuzma}, {Da Costa}, {Keller}, \&
  {Maunder}}]{kuzma2015}
{Kuzma}, P.~B., {Da Costa}, G.~S., {Keller}, S.~C., \& {Maunder}, E. 2015,
  \mnras, 446, 3297

\bibitem[{{Lane} {et~al.}(2019){Lane}, {Navarro}, {Fattahi}, {Oman}, \&
  {Bovy}}]{lane2019}
{Lane}, J. M.~M., {Navarro}, J.~F., {Fattahi}, A., {Oman}, K.~A., \& {Bovy}, J.
  2019, arXiv e-prints, arXiv:1905.12633
\bibitem[Leung \& Bovy(2019)]{leung} Leung, H.~W., \& Bovy, J.\ 2019, arXiv e-prints, arXiv:1902.08634

\bibitem[{{Miyamoto} \& {Nagai}(1975)}]{mn1975}
{Miyamoto}, M., \& {Nagai}, R. 1975, \pasj, 27, 533

\bibitem[{{Navarro} {et~al.}(1997){Navarro}, {Frenk}, \& {White}}]{navarro1997}
{Navarro}, J.~F., {Frenk}, C.~S., \& {White}, S.~D.~M. 1997, \apj, 490, 493

\bibitem[{{Odenkirchen} {et~al.}(2009){Odenkirchen}, {Grebel}, {Kayser}, {Rix},
  \& {Dehnen}}]{odenkirchen2009}
{Odenkirchen}, M., {Grebel}, E.~K., {Kayser}, A., {Rix}, H.-W., \& {Dehnen}, W.
  2009, \aj, 137, 3378

\bibitem[Price-Whelan et al.(2016)]{adrian}  Price-Whelan, A.~M., B.~Sesar, K.~V.~Johnston, and H.-W.~Rix, 2016, \apj,  824, 104
\bibitem[{{Price-Whelan}(2017)}]{gala}
{Price-Whelan}, A.~M. 2017, The Journal of Open Source Software, 2, 388
\bibitem[Price-Whelan, \& Bonaca(2018)]{price18} Price-Whelan, A.~M., \& Bonaca, A.\ 2018, \apjl, 863, L20
\bibitem[Price-Whelan et al.(2019)]{price19} Price-Whelan, A.~M., Mateu, C., Iorio, G., et al.\ 2019, arXiv e-prints, arXiv:1910.00595


\bibitem[{{Sanders} \& {Binney}(2013)}]{sanders2013}
{Sanders}, J.~L., \& {Binney}, J. 2013, \mnras, 433, 1813
\bibitem[Schlafly et al.(2014)]{schlafly} Schlafly, E.~F., Green, G., Finkbeiner, D.~P., et al.\ 2014, \apj, 789, 15 
\bibitem[Schlegel et al.(1998)]{schlegel} Schlegel, D.~J., Finkbeiner, D.~P., \& Davis, M.\ 1998, \apj, 500, 525 
\bibitem[Sch{\"o}nrich et al.(2019)]{schonrich} Sch{\"o}nrich, R., McMillan, P., \& Eyer, L.\ 2019, \mnras, 487, 3568
\bibitem[Szentgyorgyi et al.(2011)]{saint} Szentgyorgyi, A., Furesz, G., Cheimets, P., et al.\ 2011, \pasp, 123, 1188

\bibitem[Sesar et al.(2015)]{sesar2015} Sesar, B., Bovy, J., Bernard, E.~J., et al.\ 2015, \apj, 809, 59 

\bibitem[Sesar et al.(2016)]{sesar2016} Sesar, B.,  Price-Whelan, A.~M., Cohen, J.~G.  et al.\ 2016, \apjl, 816, 4 
\bibitem[Strader et al.(2011)]{strader} Strader, J., Caldwell, N., \& Seth, A.~C.\ 2011, \aj, 142, 8

\bibitem[Walker et al.(2009)]{walker09} Walker, Matthew G., Mateo, M. et al. \ 2009, \aj, 137, 3109

\bibitem[{Walt {et~al.}(2011)Walt, Colbert, \& Varoquaux}]{numpy}
Walt, S. v.~d., Colbert, S.~C., \& Varoquaux, G. 2011, Computing in Science and
  Eng., 13, 22.
\newblock \url{http://dx.doi.org/10.1109/MCSE.2011.37}


\end{thebibliography}

\end{document}